\documentclass[letterpaper,12pt]{article} 

\usepackage{amsmath}
\usepackage{amssymb}
\usepackage[round]{natbib}
\usepackage[dvips]{epsfig}
\usepackage{dcolumn}
\usepackage{enumerate}

\usepackage{hhline}
\usepackage{dsfont}
\usepackage{afterpage}
\usepackage{arydshln}
\usepackage{graphicx}
\usepackage{color}
\usepackage[usenames,dvipsnames]{xcolor}
\usepackage{rotating}
\usepackage[breaklinks]{hyperref}
\usepackage{breakurl} 
\usepackage{xr}
\usepackage[percent]{overpic}
\usepackage{subfig}
\usepackage[]{caption}
\usepackage{placeins}
\usepackage{algorithmicx}
\usepackage[noend]{algpseudocode}
\usepackage{algorithm}
\usepackage{diagbox}
\usepackage{graphicx}
\usepackage{wrapfig}
\usepackage{lscape}
\input epsf
\usepackage{fontenc}
\usepackage{setspace}
\usepackage{bm}
\usepackage{slashbox}
\usepackage{lscape}
\usepackage{breakurl} 
\usepackage{multirow}
\usepackage{eurosym}
\usepackage{titlesec}

\usepackage{float} 
\usepackage{booktabs} 
\usepackage{graphicx} 
\usepackage[margin=1cm]{caption} 
\usepackage{mathtools}
\usepackage{enumitem}

\def \avec {\text{\boldmath$a$}}

\def \dvec {\text{\boldmath$d$}}

\def \gvec {\text{\boldmath$g$}}    
    \def \mH {\text{\boldmath$H$}}

\def \uvec {\text{\boldmath$u$}}    
    
\def \wvec {\text{\boldmath$w$}}    
\def \xvec {\text{\boldmath$x$}}    
\def \yvec {\text{\boldmath$y$}}    
\def \zvec {\text{\boldmath$z$}}

\def \betavec         {\text{\boldmath$\beta$}}

\def \deltavec        {\text{\boldmath$\delta$}}
\def \epsilonvec      {\text{\boldmath$\epsilon$}}

\def \etavec          {\text{\boldmath$\eta$}}
\def \thetavec        {\text{\boldmath$\theta$}}
\def \varthetavec     {\text{\boldmath$\vartheta$}}

\def \lambdavec       {\text{\boldmath$\lambda$}}
\def \muvec           {\text{\boldmath$\mu$}}

\def \upsilonvec      {\text{\boldmath$\upsilon$}}

\def \psivec          {\text{\boldmath$\psi$}}

\def\wZ{\widetilde{Z}}
\def\bwZ{\widetilde{\bm{Z}}}

\usepackage[margin=1in]{geometry}
\titlespacing*\section{0pt}{0pt plus 2pt minus 2pt}{0pt plus 2pt minus 2pt}
\titlespacing*\subsection{0pt}{0pt plus 2pt minus 2pt}{0pt plus 2pt minus 2pt}
\titlespacing*\subsubsection{0pt}{0pt plus 2pt minus 2pt}{0pt plus 2pt minus 2pt}

\newcounter{mynotation}

\usepackage{paralist}

\renewenvironment{itemize}[1]{\begin{compactitem}#1}{\end{compactitem}}

\makeatletter
\def\@seccntformat#1{\@ifundefined{#1@cntformat}%
	{\csname the#1\endcsname\quad}  % default
	{\csname #1@cntformat\endcsname}% enable individual control
}
\let\oldappendix\appendix %% save current definition of \appendix
\renewcommand\appendix{%
	\oldappendix
	\newcommand{\section@cntformat}{\appendixname~\thesection\quad}
}
\makeatother

\renewcommand{\baselinestretch}{1.8}
\begin{document}
\pagestyle{empty}
\begin{titlepage}
\title{Regression Copulas for Multivariate Responses}
\author{Nadja Klein, Michael Stanley Smith, David J. Nott and Ryan Chisholm}
\date{}
\maketitle
\noindent
{\small Nadja Klein is Professor of Uncertainty Quantification and Statistical Learning at Research Center  Trustworthy Data Science and Security (UA Ruhr) and Department of Statistics (Technische Universit\"at Dortmund); Michael Stanley Smith is Professor of Management (Econometrics) at Melbourne Business School, University of Melbourne; David J. Nott is Associate Professor of Statistics and Applied Probability at National University of Singapore; Ryan Chisholm is Associate Professor of Biological
Sciences at National University of Singapore. 
Correspondence should be directed to Nadja Klein at 
{\tt nadja.klein@tu-dortmund.de}.\\

\noindent \textbf{Acknowledgments:} Nadja Klein acknowledges support through the Emmy Noether grant KL 3037/1-1 of the German research foundation (DFG). David Nott's research was supported by the Ministry of Education, Singapore, under the Academic Research Fund Tier 2 (MOE-T2EP20123-0009). David Nott
is affiliated with the Institute of Operations Research and Analytics, National University of Singapore.}\\

\newpage
\begin{center}
	\mbox{}\vspace{2cm}\\
	{\LARGE \title{Regression Copulas for Multivariate Responses}
	}\\
\end{center}
\vspace{-1pt}
\onehalfspacing
\noindent

\date{}
\maketitle
\begin{abstract}
	\noindent
We propose a novel distributional regression model for a multivariate response vector based on a copula process
over the covariate space. It uses the implicit copula of a Gaussian multivariate regression, which we call a ``regression copula''.
To allow for large covariate vectors their coefficients are regularized 
using a novel multivariate extension of the horseshoe prior. Bayesian inference and
distributional predictions are evaluated using efficient
variational inference methods, allowing application to large datasets.
An advantage of the approach 
is that the marginal distributions of the response vector can be estimated
separately and accurately, resulting in predictive distributions that are marginally-calibrated.
Two substantive applications of the methodology highlight its efficacy 
in multivariate modeling. 
The first is the econometric modeling and prediction of half-hourly
regional Australian electricity prices. Here, our approach produces more accurate
distributional forecasts than leading benchmark methods. 
The second is the evaluation of multivariate posteriors in 
likelihood-free inference (LFI) of a model for tree species abundance  
data, extending a previous univariate regression copula LFI method.  In both applications, we demonstrate that our new approach exhibits a desirable marginal calibration property.

\end{abstract}
\vspace{20pt}

\noindent
{\bf Keywords}:
 Copula Process; Distributional Regression; Implicit Copula; Likelihood-Free Inference; Semiparametric Multivariate Regression; Variational Inference
\end{titlepage}

\renewcommand{\baselinestretch}{1.8}
\newpage
\pagestyle{plain}
\setcounter{equation}{0}
\renewcommand{\theequation}{\arabic{equation}}
\section{Introduction}%\label{sec:intro}
Distributional regression, which extends traditional mean based regression models to allow the entire distribution
to be a function of a covariate vector, is increasingly popular.
%; see~\cite{Kle2024} for an overview. 
For a univariate response,
there are many approaches including (but not restricted to) GAMLSS models~\citep{rigby2005}, conditional transformation models~\citep{HotKneBuh2014}, mixtures of experts models~\citep{JordanJacobs1994}, quantile regression~\citep{KoeBas1978} 
and regression copula models~\citep{KleSmi2019,smithklein2020}. 
However, effective distributional regression methods
for a multivariate response vector are much more limited. 
%For example,~\cite{KleHotBarKne2022} extend the transformation model of~\cite{HotKneBuh2014} to this case, but for only a single covariate, while
%\citet{KocKle2023} extend GAMLSS to multivariate responses.
In this paper we 
extend the regression copula model to the multivariate case, and 
demonstrate its advantages in two
challenging applications, one from econometric modeling and another 
from likelihood free inference in ecology. 

\cite{KleSmi2019} combine a copula process over the covariate space with an arbitrary marginal for a univariate response to define a
distributional regression. Distributional predictions are given by the Bayesian predictive distribution from the copula model.
The copula used is the implicit copula of the response values (called pseudo responses hereafter) from a Bayesian regularized Gaussian regression model with
its regression coefficients marginalized out. 
The authors call this a ``regression copula'' because it is a function of the covariates, a term that we also employ here. To extend this approach we 
construct an implicit copula from a Bayesian multivariate regularized Gaussian regression model with the
regression coefficients marginalized out. This presents a number of challenges. First, 
an effective regularization prior is required for the regression coefficients, and for this 
we propose a novel multivariate extension 
of the horseshoe prior~\citep{CarPolSco2009,CarPol2010}. This conjugate prior is scaled by the correlation matrix of the
regression disturbance in the 
same way as the g-priors of~\cite{brown1998} and~\cite{SmiKoh2000}. We show
that the marginal
priors for the regression coefficients of each 
equation in the multivariate regression are standard horseshoe priors. The second challenge is the 
selection of a prior
for the cross-equation correlation matrix, and we consider two choices. The first is an order-invariant prior suggested by~\cite{ArcRei2021}, and the second is based on a factor model as proposed by~\cite{murray2013} and allows application of our model to high-dimensional response
vectors.

A third challenge is that, unlike 
\cite{KleSmi2019} who use MCMC to evaluate the posterior
of their univariate distributional regression, application of MCMC is 
difficult here because the posterior is both more complex and of a much higher dimension. 
Therefore, we use an efficient
variational inference (VI) method with a parsimonious
Gaussian approximation and stochastic gradient optimization methods that employ efficient reparameterized gradients~\citep{OngNotSmi2018}. This is faster and more
robust than MCMC. A final challenge is that the joint posterior is computationally intractable.
Therefore we propose using an augmented posterior density that is fast to compute as the
the target of our variational 
optimization. The end result is a method that can be employed with large datasets in high dimensions and with 
sizeable covariate vectors.

A major advantage of our method is that the user has complete control over 
the choice of marginal distribution for each response variable. Moreover, the predictive distributions are marginally-calibrated, which is where their long run average matches this marginal~\citep{Gneetal2007}. These advantages, and the efficacy of our approach, are shown in two demanding applications. The first is the modeling and forecasting of intraday Australian regional electricity prices using a large dataset. 
Here, we show that our approach can account for the complex non-Gaussian distributions of electricity prices, the nonlinear
effect of regional demands on each regional price, and the strong inter-regional dependence in prices. In an extensive forecasting study we show that the marginally-calibrated distributional forecasts are more accurate than a range of benchmark methods. 

The second application is to likelihood-free inference (LFI)
of an ecological model of tree species abundance census data.  In LFI, distributional regression
methods can be used to approximate the posterior distribution when the likelihood is intractable or inference
is based only on summary statistics. 
In this ecological example our multivariate distributional regression method extends a previous univariate LFI regression copula approach by approximating the full joint posterior distribution while maintaining a desirable marginal calibration property.
An analysis of tree species abundance fluctuations also demonstrates
the benefits of using LFI methods to deal
with model misspecification through summary statistic based inference.

% We demonstrate that 
%our multivariate distributional regression method improves on LFI
%compared to the univariate special case 
%considered in~\cite{KleNotSmi2021} which
%cannot estimate the full joint posterior density, but only approximates its marginal components. An analysis of species abundance fluctuations also demonstrate
%the benefits of using LFI methods to deal
%with model misspecification through summary statistic based inference,
%and of joint posterior estimation, are demonstrated in the estimation
%of the variance of species abundance fluctuations.

The rest of the paper is organized as follows. Section~\ref{sec:MVC} shows how to
specify a copula process model for distributional regression. 
Section~\ref{sec:copprocess} specifies the implicit copula of the pseudo responses
from a regularized multivariate regression. This is a Bayesian formulation using 
carefully constructed priors. Section~\ref{sec:VI} describes the VI method for 
estimation and distributional prediction, while Section~\ref{sec:electricity}
contains the econometric application. Section~\ref{sec:ecology} applies our distributional
regression to multivariate posterior estimation in LFI for the ecological model, while Section~\ref{sec:discussion} concludes.  

%In the trivial case where the multivariate response has only a single element, our
%extended regression copula model is equivalent to that in~\cite{KleSmi2019}. But
\setlength{\abovedisplayskip}{1pt}
\setlength{\belowdisplayskip}{1pt}
\section{Copula Model for Multivariate Regression}\label{sec:MVC}
We first outline a regression model  for a vector-valued response $\bm{Y}_i=(Y_{i,1},\ldots,Y_{i,p})^\top$
that is based on a copula process. 
It is a distributional regression model, %\citep[see e.g.,][for a review]{Kle2024}
which is where the
entire predictive distribution of $\bm{Y}_i$ is a function of a covariate vector $\xvec_i=(x_{i,1},\ldots,x_{i,q})^\top$.
The employed copula process  will be defined  in Section~\ref{sec:copprocess}.

\subsection{Copula process model}
Denote $n$ realizations on each continuous-valued dependent variable as $\bm{Y}_{1:n,j}=(Y_{1,j},\ldots,Y_{n,j})^\top$ for $j=1,\ldots,p$, 
and corresponding covariate vectors $\xvec_{1:n}=\{\xvec_1,\ldots,\xvec_n\}$ that are common across response components $j$. 
Then, we define the distribution of 
 $\bm{Y}_{1:n}=(\bm{Y}_{1:n,1}^\top,\ldots,\bm{Y}_{1:n,p}^\top)^\top$ (that is, the dependent variables stacked first by realization and then by variable), conditional on $\xvec_{1:n}$, using a copula model with distribution function
%
%Let $\varthetavec$ denote copula parameters that are unaffected by $n$, 
%then a copula model is adopted for $\bm{Y}_{1:n}|\xvec_{1:n},\varthetavec$ with distribution function 
%representation
\begin{equation}
F_{Y_{1:n}}(\yvec_{1:n}|\xvec_{1:n},\varthetavec)=C_{1:n}\left(\uvec_{1:n,1}^\top,\ldots,\uvec_{1:n,p}^\top; \xvec_{1:n},\varthetavec\right)\,,\;\mbox{ for }n\geq 2\,,\;p\geq 1.\label{eq:copmodcdf}
\end{equation}
Here, $C_{1:n}$ is a copula function\footnote{For a vector  $\uvec=(u_1,\ldots,u_m)^\top$, we denote copula functions interchangeably as $C(u_1,u_2,\ldots,u_m)$ and 
	$C(\uvec)$, and copula densities as $c(u_1,\ldots,u_m)$ and $c(\uvec)$.} 
on the unit cube $[0,1]^{np}$ with uniform margins. 
The arguments of the copula function are
$\uvec_{1:n,j}^\top=(u_{1,j},\ldots,u_{n,j})$ with $u_{i,j}=F_{Y_{i,j}}(y_{i,j}|\xvec_i)$, where
$F_{Y_{i,j}}$ is the cumulative distribution function of $Y_{i,j}|\xvec_i$, which is the margin of $F_{Y_{1:n}}$.
In this paper the copula
is a function
of $\xvec_{1:n}$ with parameters $\varthetavec$ that do not vary with $n$, so that it is a
``copula process''~\citep{Wilson2010} on the covariate space. 

If  $\uvec_{1:n}=(\uvec_{1:n,1}^\top,\ldots,\uvec_{1:n,p}^\top)^\top$, then
differentiating~\eqref{eq:copmodcdf} with respect to $\yvec$ gives
the density
\begin{equation}
f_{Y_{1:n}}(\yvec_{1:n}|\xvec_{1:n},\varthetavec)=c_{1:n}\left(\uvec_{1:n}; \xvec_{1:n},\varthetavec\right)\prod_{i=1}^n \prod_{j=1}^p f_{Y_{i,j}}(y_{i,j}|\xvec_i)\,,\label{eq:copmodpdf}
\end{equation}
with $f_{Y_{i,j}}(y|\xvec_i)=\frac{d}{d y } F_{Y_{i,j}}(y|\xvec_i)$ and $c_{1:n}(\uvec;\xvec_{1:n},\varthetavec)=\frac{\partial^{np}}{\partial \uvec}C_{1:n}(\uvec;\xvec_{1:n},\varthetavec)$ is commonly called the ``copula density''. 

One advantage of the model at~\eqref{eq:copmodcdf} is that if $F_{Y_{i,j}}=G_j$ is 
assumed invariant with respect to index $i$ and covariates $\xvec_i$, then the margins $G_1,\ldots,G_p$ can be estimated 
using nonparametric or other 
flexible estimators, as we do here.
%; see the discussion in~\cite{KleNotSmi2021}.
Because the copula $C_{1:n}$ is a function of the covariates $\xvec_{1:n}$, \cite{smithklein2020} call such a copula process a
``regression copula''. We stress that this model should not be confused with a 
copula model
with marginals that are regressions and a parametric copula
that is either independent of the covariates \citep[as
in][]{PitChaKoh2006,song2009} or also covariate-dependent \citep[as in][]{AcaCraYao2013}.

\subsection{Multivariate distributional regression}
The copula model at~\eqref{eq:copmodcdf} and~\eqref{eq:copmodpdf} specifies a distributional multivariate regression
model, even when $G_1,\ldots,G_p$ are not functions of $\xvec_i$. To see why, consider the prediction of a new realization of the response
vector
$\bm{Y}_{n+1}=(Y_{n+1,1},\ldots,Y_{n+1,p})^\top$. The predictive density,\footnote{While we denote the copula process density and the likelihood functions with subscripts to help distinguish them, we adopt the usual Bayesian style of  denoting posteriors and predictive densities with an unsubscripted generic $p$.}
	conditional on existing observations $\bm{Y}_{1:n}=\bm{y}_{1:n}$,  
 covariates $\xvec_{1:n+1}$ and parameters $\varthetavec$, is
\begin{eqnarray}
p(\yvec_{n+1}|\yvec_{1:n},\xvec_{1:n+1},\varthetavec) &= &\frac{f_{Y_{1:n+1}}(\yvec_{1:n+1}|\xvec_{1:n+1},\varthetavec)}{f_{Y_{1:n}}(\yvec_{1:n}|\xvec_{1:n},\varthetavec)}
		=\frac{c_{1:n+1}(\uvec_{1:n+1};\xvec_{1:n+1},\varthetavec)}
		{c_{1:n}(\uvec_{1:n+1};\xvec_{1:n},\varthetavec)}\prod_{j=1}^p g_j(y_{n+1,j}) \nonumber\\
	&=&	p(\uvec_{n+1}|\uvec_{1:n},\xvec_{1:n+1},\varthetavec)\prod_{j=1}^p g_{j}(y_{n+1,j})\,, \label{eq:predy}
\end{eqnarray}
where  the marginal density of $Y_{i,j}$ is $f_{Y_{i,j}}=g_j=\frac{dG_j}{dy_j}$. Thus, the entire predictive density is
a function $\xvec_{n+1}$ (and also $\xvec_{1:n}$) through the first term in~\eqref{eq:predy} that is due to the copula process. 
A Bayesian method for 
computing~\eqref{eq:predy} efficiently for the copula process proposed in this paper is outlined in 
Section~\ref{sec:predinf}. 

%\subsection{Marginal calibration}

%The predictive distributions obtained from a distributional regression model
%can be thought of as probabilistic forecasts.  
%One criterion for evaluating probabilistic forecasts is ``calibration", which
%is concerned with statistical consistency of the observations with the
%forecasts \citep{Gneetal2007}.  \cite{KleNotSmi2021} considered the 
%univariate version of the model proposed here, and 
%demonstrate that their copula regression approach performs better
%than alternative distributional regression methods 
%in terms of ``marginal calibration".  \cite{Gneetal2007} formulate
%the notion of marginal calibration by considering a sequence
%of forecast distributions with corresponding distributions for
%a true state of nature, with marginal calibration defined as equality of
%the average forecast distribution and nature's average distribution 
%in the limit of an increasing number of forecasts.  
%Another formulation is given in 
%\cite{gneiting2013}, in terms of equality between the expected forecast
%distribution and the distribution of the 
%true state of nature, without considering any long run 
%sequence of forecasts.  We will further discuss marginal calibration 
%in the application in Section \ref{sec:ecology}, where marginal calibration can be used to check the accuracy of
%posterior approximations in likelihood-free inference.  %We demonstrate
%that our proposed multivariate regression copula method is marginally
%calibrated, similar to the method of \cite{KleNotSmi2021}.

\section{Implicit Copula Process}\label{sec:copprocess}
The choice of copula at~\eqref{eq:copmodcdf} is key to our multivariate response regression model, for which we use an ``implicit copula''
with parameters $\varthetavec$.
This is the copula
of a parametric multivariate distribution constructed
by inversion of Sklar's theorem.
%; see~\cite{smith2021} for an overview of this class of copulas.
In particular, we use the implicit copula of the distribution of realizations
from a Gaussian multivariate
regression with the regression coefficients integrated out. The dependent variables of these regressions
are called ``pseudo'' responses because they are unobserved.\footnote{In this section, the pseudo response is denoted $\widetilde{Z}_{i,j}$, and should not
	be confused with the response $Y_{i,j}$ of the distributional regression model.
The pseudo responses are only introduced to enable construction of their implicit copula.}  
Our model entails a Bayesian formulation for which we propose a novel extension
of the horseshoe  prior~\citep{CarPolSco2009,CarPol2010} for the entire vector of regression coefficients.
%Unlike in the previous section, we do not adopt
%observational subscripts (eg.
%$\bwZ_{1:n,j}$ or $\xvec_{1:n}$) in order to make the expressions more succinct.

\subsection{Multivariate regression model}
Let $\bwZ_{1:n,j}=(\wZ_{1,j},\dots, \wZ_{n,j})^\top$ 
be a vector of $n$ realizations on a pseudo variable.
For each variable $j=1,\ldots,p$, we consider a linear regression with the same $q$
covariates, given by 
\begin{equation}
	\bwZ_{1:n,j} = F\betavec_j+\epsilonvec_j\,, \mbox{ for }j=1,\ldots,p\,,\label{eq:linmod}
\end{equation}
where $F=\left[\xvec_{1}|\cdots|\xvec_{n}\right]^\top$ is an $n \times q$ 
design matrix and $\betavec_j=(\beta_{1,j},\dots, \beta_{q,j})^\top$ are the coefficients. The regression has Gaussian 
distributed errors $\epsilonvec_j=(\epsilon_{1,j},\ldots,\epsilon_{n,j})^\top$ that are correlated 
across equations but not observations,
with $E(\epsilon_{i,j}\epsilon_{i',l})=\sigma_{j,l}$ if $i=i'$, and zero otherwise.

We stack the pseudo-responses and errors in the same order as  $\bm{Y}_{1:n}$ in the previous section,
so that 
$\bwZ_{1:n}=(\bwZ_{1:n,1}^{\top},\dots, \bwZ_{1:n,p}^{\top})^\top$ and 
$\epsilonvec=(\epsilonvec_1^{\top},\dots, \epsilonvec_p^{\top})^\top$. Then,
if $\betavec=(\betavec_{1}^\top,\dots, \betavec_p^{\top})^\top$, and
$X= I_p \otimes F$ (where $\otimes$ denotes the Kronecker product and $I_p$ 
a $p\times p$ identity matrix),
the system of $p$
regressions can be written in stacked form as
\begin{equation}
	\bwZ_{1:n} = X\betavec+\epsilonvec\,,\;\mbox{ with }\epsilonvec| \Sigma \sim N(\bm{0},\Sigma\otimes I_n)\,,\label{eq:sur}
\end{equation}
with $\Sigma=\{\sigma_{j,l}\}_{j=1:p,\,l=1:p}$ a $p\times p$ covariance matrix.
This model is the ``seemingly unrelated regression'' (SUR)  of~\cite{zellner1962} with common covariates for each of the $p$ 
regression equations. 

\subsection{Extended horseshoe prior}
Regularization of $\betavec$ is advantageous when either 
$q$ or $p$ is large, including when $F$ contains function basis terms as in Section~\ref{sec:electricity}.  This is achieved in a Bayesian analysis
through its prior,
%; for example, 
%popular slab-and-spike priors to do so 
%are given in~\cite{brown1998} and~\cite{SmiKoh2000}
and here we propose a novel  prior for $\betavec$ that extends
the horseshoe prior to the multivariate  response regression case.
To specify this prior, we first define two matrix operators as follows.
Let bdiag be a matrix operator with $D=\mbox{bdiag}(D_1,\ldots,D_p)$ the $pq \times pq$ block diagonal matrix with $q\times q$ blocks $D_j$ each of the same dimension.   
If each
block $D_j$ is positive definite with upper triangular Cholesky factorization $D_j=U _j^\top U _j$, then define the operator ``$\star$'' so that 
\[
\Sigma \star D = \{\sigma_{jl}U _j^\top U _l\}_{j=1:p,l=1:p}=U ^\top (\Sigma \otimes I_q )U\,,
\]
for
$U =\mbox{bdiag}(U _1,\ldots,U _p)$.  With this notation, the conjugate prior
we use is
\begin{equation}
	\betavec | \Sigma,\thetavec \sim N\left(\bm{0},\Sigma \star P(\thetavec)^{-1}\right)\,,\label{eq:betaprior}
\end{equation}
where $P(\thetavec)=\mbox{bdiag}(P_1(\thetavec_1),\ldots,P_p(\thetavec_p))$ and $\thetavec = (\thetavec_1^\top, \dots, \thetavec_p^\top)^\top$. The matrix
$P_j(\thetavec_j)$ is the diagonal precision matrix of 
a standard horseshoe prior for regression $j$ with hyper-parameters $\thetavec_j$, for $j=1,\ldots,p$. Writing
the Cholesky factors of $P_j^{-1}$ and $P^{-1}$ as $P_j^{-1/2}$ and $P^{-1/2}$, respectively, 
the $pq \times pq$ prior covariance matrix at~\eqref{eq:betaprior} can be expressed as
\[
\Sigma \star P(\thetavec)^{-1} = 
\{\sigma_{jl}P_j(\thetavec_j)^{-1/2}P_l(\thetavec_l)^{-1/2}\}_{j=1:p,l=1:p}
=P(\thetavec)^{-1/2}(\Sigma \otimes I_q)P(\thetavec)^{-1/2}\,.
\]

We make three observations concerning this prior. First, 
the prior covariance is scaled by $\Sigma$ in the same manner
as the $g$-prior for a SUR model given in~\cite{brown1998} and~\cite{SmiKoh2000}.
Second, the marginal prior is $\betavec_j|\Sigma,\thetavec\sim N(\bm{0},\sigma_{jj}P_j^{-1}(\thetavec_j))$, 
which is a standard horseshoe regularization prior
for regression equation $j$ at~\eqref{eq:linmod}. Third, the prior is a conjugate but dependent
prior for $\betavec_1,\ldots,\betavec_p$ (e.g. $E(\betavec_j\betavec_l^\top)=\sigma_{jl}P_j(\thetavec_j)^{-1/2}P_l(\thetavec_l)^{-1/2}$)
in contrast to independent horseshoe priors as in~\cite{li2019}. Adopting our prior
greatly simplifies the expression and computation of the posterior with $\betavec$ marginalized out as
in Section~\ref{sec:mregcop} below.

To finalize the regularized regression specification, as in the standard horseshoe prior we specify  $P_j(\thetavec_j)^{-1}=\mbox{diag}(\xi_{j,1}^2,\ldots,\xi_{j,q}^2)$,
with hyper-priors $\xi_{j,k}|\tau_j\sim \mbox{Half-Cauchy}(0,\tau_j)\,, \mbox{ and }
\tau_j\sim \mbox{Half-Cauchy}(0,1)$.
Setting $\thetavec_j=(\xi_{j,1},\ldots,\xi_{j,p},\tau_j)^\top$, we assume these hyper-parameters
are independent across regressions $j=1,\ldots,p$ a priori.

\subsection{Regression copula derivation}\label{sec:mregcop}
Following~\cite{KleSmi2019} and~\cite{smithklein2020} we construct the implicit copula of the
distribution of $\bwZ_{1:n}|\xvec_{1:n},\Sigma,\thetavec$
with $\betavec$
integrated out. (Notice that if $\betavec$ was not integrated out, then from~\eqref{eq:linmod} the implicit copula of 
$\bwZ_{1:n}|\xvec_{1:n},\Sigma,\betavec,\thetavec$ is simply the independence copula).
	Recognizing a Gaussian in $\betavec$ and applying
the Woodbury formula  (see Part~A.1 of the Web Appendix) gives
$\bwZ_{1:n}|\xvec_{1:n},\Sigma,\thetavec \sim N(\bm{0},V)$ with
\begin{equation*}
	V=(\Sigma \otimes I_n) + X(\Sigma \star P(\thetavec)^{-1})X^\top\,,
\end{equation*}
%\begin{eqnarray*}
%	V &= &\left( (\Sigma \otimes I_n)^{-1} -(\Sigma \otimes I_n)^{-1} X
%	\left[X^\top (\Sigma \otimes I_n)^{-1} X + (\Sigma \star P(\thetavec)^{-1})^{-1} 
%	\right]^{-1} X^\top(\Sigma \otimes I_n)^{-1} \right)^{-1}\\
%	&= & (\Sigma \otimes I_n) + X(\Sigma \star P(\thetavec)^{-1})X^\top\,,
%\end{eqnarray*}
 In our model
$X=(I_p \otimes F)$, and with some straightforward matrix algebra, the second term can be simplified further as
\begin{eqnarray*}
X(\Sigma \star P(\thetavec)^{-1})X^\top &= &(I_p\otimes F)(\Sigma \star P(\thetavec)^{-1})(I_p \otimes F^\top)\\
&= &\left\{ \sigma_{lj} FP_l(\thetavec_l)^{-1/2}P_j(\thetavec_j)^{-1/2}F^\top\right\}_{l=1:p,j=1:p}
=\widetilde{F}(\Sigma \otimes I_q)\widetilde{F}^\top\,,
\end{eqnarray*}
with
$\widetilde{F}=\mbox{bdiag}(FP_1(\thetavec_1)^{-1/2},\cdots,FP_p(\thetavec_p)^{-1/2})$.

Write the diagonal matrix $S=\mbox{diag}(V)^{-1/2}$, then 
the implicit copula of a Gaussian distribution is a
Gaussian copula~\citep{Song2000}
with a parameter matrix given by the correlation matrix $R=SVS$.
In Part~A.2 of the Web Appendix we establish the following two results. 
First,  $R$ is a function of $\Sigma$ only through the term
$\mbox{diag}(\Sigma)^{-1/2}\Sigma\, \mbox{diag}(\Sigma)^{-1/2}$. Therefore, 
without loss of generality, we set $\sigma_{jj}=1$ for all $j$, so $\Sigma$
is strictly a correlation matrix that is identified in the copula. The second result is that 
$S=\mbox{bdiag}(S_1,\ldots,S_p)$, with
\[
S_j=\mbox{diag}\left( (1+\xvec_1^\top P_j(\thetavec_j)^{-1} \xvec_1)^{-1/2},\ldots,(1+\xvec_n^\top P_j(\thetavec_j)^{-1}\xvec_n)^{-1/2}
\right)\,.
\]
Hence, $S$ is a diagonal matrix which can be computed in $O(npq)$ operations because evaluating $\xvec_i^\top P_j(\thetavec_j)^{-1} \xvec_i$ is $O(q)$.
 
%The implicit copula of a Gaussian distribution is a
%Gaussian copula~\citep{Song2000}
%with a parameter matrix given by the correlation matrix of $V$.  
%Write the $np\times np$ diagonal matrix $S=\mbox{diag}(V)^{-1/2}$, which is
%given by $S=\mbox{bdiag}(S_1,\ldots,S_p)$ with the matrices
%\begin{equation}
%S_j=\left(\sigma_{jj}\mbox{diag}(I_n+FP_j(\thetavec_j)^{-1} F^\top)\right)^{-1/2}\,,\mbox{ for }j=1,\ldots,p\,,
%\end{equation} 
%being diagonal.
%Define $\bm{Z}=S\bwZ$ as the ``normalized'' pseudo-response vector, which has
%distribution
%$\bm{Z}|\Sigma,\thetavec \sim N\left(\bm{0},R(\Sigma,\thetavec,\xvec_{1:n})\right)$, with $R(\Sigma,\thetavec,\xvec_{1:n})=SV(\Sigma,\thetavec,\xvec_{1:n})S$ a correlation matrix. This is also 
%only a function of $\mbox{diag}(\Sigma)^{-1/2}\Sigma\, \mbox{diag}(\Sigma)^{-1/2}$,
%so that
%without loss of generality we can set $\sigma_{jj}=1$ for all $j$ (and $\Sigma$
%is therefore strictly a correlation matrix). 
%Because $F=[\xvec_1|\cdots|\xvec_n]^\top$, 
%\[
%S_j=\mbox{diag}\left( (1+\xvec_1^\top P_j(\thetavec_j)^{-1} \xvec_1)^{-1/2},\ldots,(1+\xvec_n^\top P_j(\thetavec_j)^{-1}\xvec_n)^{-1/2}
%\right)\,,
%\]
%evaluation of 
%which is an $O(nq)$ computation. 

Denoting $R$ as a function of $\{\Sigma,\thetavec,\xvec_{1:n}\}$, the $np$-dimensional copula density at~\eqref{eq:copmodpdf} is
the well-known Gaussian copula density
%~\citep{Song2000} 
given by 
\begin{equation}
	c_{1:n}(\uvec_{1:n};\xvec_{1:n},\varthetavec)  = \frac{\phi_{np}\left(\zvec_{1:n};\bm{0},R(\Sigma,\thetavec,\xvec_{1:n})\right)}{\prod_{i=1}^n \prod_{j=1}^p \phi_1(z_{i,j})},  \label{eq:gcop}
\end{equation}
where 
%$\uvec_{1:n}=(\uvec_{1:n,1}^{\top},\dots,\uvec_{1:n,p}^{\top})^\top$, $\uvec_{1:n,j}=(u_{1,j},\dots, u_{n,j})^\top$, 
$\zvec_{1:n}=(\zvec_{1:n,1}^{\top},\dots,\zvec_{1:n,p}^{\top})^\top$, $\zvec_{1:n,j}=(z_{1,j},\dots, z_{n,j})^\top$ and $z_{i,j}=\Phi_1^{-1}(u_{i,j})$.
Here,
$\phi_d(\cdot;\muvec,\Omega)$ and $\Phi_d(\cdot;\muvec,\Omega)$ denote the density and 
distribution functions of a $N_d(\muvec,\Omega)$ distribution, respectively, 
and we write simply $\phi_1(\cdot)$ and $\Phi_1(\cdot)$
when $d=1$, $\muvec=0$ and $\Omega=1$. 
The copula parameters are $\varthetavec=\{\Sigma,\thetavec\}$. 
This is a Gaussian copula process on the covariate space, because the correlation matrix 
$R$ in~\eqref{eq:gcop} is a function of the covariate values $\xvec_{1:n}$. When $p=1$
it simplifies to the copula process for a univariate response  given in~\cite{KleNotSmi2021}. 

%We note that conditional
%on $\{\xvec_{1:n},\thetavec,\Sigma\}$, 

The random vector
$\bm{Z}_{1:n}=S\bwZ_{1:n}\sim N_{np}(\bm{0},R)$, from which $\zvec_{1:n}$ in~\eqref{eq:gcop} is a drawn. These are dependent upon
$\{\xvec_{1:n},\thetavec,\Sigma\}$, which is an observation that 
is important when considering estimation and inference in Section~\ref{sec:VI}.

%
%Finally,
%the $np$-dimensional Gaussian copula density is therefore
%\begin{equation}
%c_{1:n}(\uvec;R(\Sigma,\thetavec,\xvec_{1:n}))  = \frac{\phi_{np}\left(\zvec;\bm{0},R(\Sigma,\thetavec,\xvec_{1:n})\right)}{\prod_{i=1}^n \prod_{j=1}^p \phi_1(z_{ij};0,1)},  \label{eq:gcop}
%\end{equation}
%where $\uvec=(\uvec_1^{\top},\dots,\uvec_p^{\top})^\top$, $\uvec_j=(u_{1,j},\dots, u_{n,j})^\top$, $\zvec=(\zvec_1^{\top},\dots,\zvec_p^{\top})^\top$, $\zvec_j=(z_{1,j},\dots, z_{n,j})^\top$ and $z_{i,j}=\Phi_1^{-1}(u_{i,j})$.
%In this specification,
%$\phi_d(\cdot;\muvec,\Omega)$ and $\Phi_d(\cdot;\muvec,\Omega)$ denote the density and 
%distribution functions of a $N_d(\muvec,\Omega)$ distribution,
%which are simplified to $\phi_1(\cdot)$ and $\Phi_1(\cdot)$
%when $\muvec=0$ and $\Omega=1$.

\subsection{Priors for $\Sigma$}
There are a range of different parameterizations and priors
 for correlation matrices~\citep{ghosh2021}. 
In our model, $\Sigma$ captures cross-sectional dependence, and we employ
two priors suitable for this case. The first (labeled ``Prior~1'') 
is based on a re-parameterization of $\Sigma$ to a vector $\upsilonvec\in \mathbb{R}^{p(p-1)/2}$
proposed by~\cite{ArcRei2021}. This prior is order invariant, and the elements of $\upsilonvec$ are on 
the same scale, which simplifies the selection of a hyperprior for them.
The second  (labeled ``Prior~2'') is based on 
a factor model as proposed by~\cite{murray2013}, which has the advantage that it is scalable with $p$. 
Appendix~\ref{app:A} details both priors and the resulting parameters $\varthetavec$. 
% Prior 1 for $\Sigma$ is hierarchical
%and incorporates a hyperparameter with a hyper-prior, but 
%in order to discuss both priors in a common framework we focus on
%the case of a fixed prior density $p(\Sigma)$ in the next section.  

%Hence, the prior 
%\begin{align*}
%\pi(l_{ii})=\pi(\exp(l_{ii}))\exp(l_{ii})=\frac{\alpha}{2\beta} \left(1+\frac{\exp(l_{ii})}{\beta} \right)^{-(\alpha+1)}\exp(l_{ii}).
%\end{align*}
%Assuming that the priors on the components of $g^0$ are independent, we get
%\begin{align*}
%\pi(g^0) &= \left(\prod_{i=2}^p \prod_{j=1}^{\min(i-1,K)} \pi(g_{ij}) \right) \prod_{k=1}^K \pi(l_{kk}) \\
%&= \left(\prod_{i=2}^p \prod_{j=1}^{\min(i-1,K)} \frac{\alpha}{2\beta}\left(1+\frac{|g_{ij}|}{\beta}\right)^{-(\alpha+1)} \right) \left(\prod_{k=1}^K \frac{\alpha}{2\beta} \left(1+\frac{\exp(l_{kk})}{\beta} \right)^{-(\alpha+1)}\exp(l_{kk}) \right)\\
%&= \left(\frac{\alpha}{2\beta}\right)^{pK-K(K-1)/2} \left(\prod_{i=2}^p \prod_{j=1}^{\min(i-1,K)} \left(1+\frac{|g_{ij}|}{\beta}\right)^{-(\alpha+1)} \right) \left(\prod_{k=1}^K \left(1+\frac{\exp(l_{kk})}{\beta} \right)^{-(\alpha+1)}\right)\exp\left(\sum_{k=1}^K l_{kk}\right).
%\end{align*}
\section{Estimation}\label{sec:VI}
\subsection{Parameter Augmentation}
From~\eqref{eq:copmodpdf} and~\eqref{eq:gcop}, the likelihood of our copula model is
\begin{equation}
	f_{Y_{1:n}}(\yvec_{1:n};\xvec_{1:n},\varthetavec)=\phi_{np}\left(\zvec_{1:n};\bm{0},R(\Sigma,\thetavec,\xvec_{1:n})\right) \prod_{i=1}^n \prod_{j=1}^p \frac{g_{j}(y_{i,j})}{\phi_1(z_{i,j})}\,.\label{eq:like1}
\end{equation}
However, because $R$ is a dense $n\times n$ matrix, evaluating the likelihood  becomes
 computationally intractable for large $n$.  Therefore, we 
 %follow  \cite{KleSmi2019} and 
 use an augmentation method
 for posterior computation that avoids evaluating $R$ directly by 
 reintroducing the parameters $\betavec$ from the model for the pseudo-responses \eqref{eq:sur} as follows. 
 
Let $\etavec=\{\betavec,\varthetavec\}$ and $J(\varthetavec,\xvec_{1:n})=\prod_{i=1}^n \prod_{j=1}^p \frac{g_{j}(y_{i,j})}{\phi_1(z_{i,j})}$ be the term arising from the transformation of variables from $\bm{Y}_{1:n}$ to $\bm{Z}_{1:n}$ \footnote{We adopt this notation
	to clarify the dependence of $\bm{Z}_{1:n}\sim N(\bm{0},R(\varthetavec,\xvec_{1:n}))$ on $\{\varthetavec,\xvec_{1:n}\}$.}. 
Then we evaluate the augmented posterior
\begin{eqnarray}
	p(\etavec|\yvec_{1:n},\xvec_{1:n}) &\propto &p(\yvec_{1:n}|\xvec_{1:n},\etavec)p(\etavec) \\
	&\propto& p(\zvec_{1:n}|\xvec_{1:n},\betavec,\varthetavec)J(\varthetavec,\xvec_{1:n})p(\betavec|\varthetavec)p(\varthetavec)  \nonumber\\
	&= &\phi_{np}(\zvec_{1:n};SX\betavec,S(\Sigma\otimes I_n)S)J(\varthetavec,\xvec_{1:n})p(\betavec|\Sigma,\thetavec)p(\varthetavec) \,. \label{eq:augpost}
\end{eqnarray}
The marginal density in $\varthetavec$ of the augmented posterior above is
the posterior obtained using the likelihood at~\eqref{eq:like1} and an identical
prior. However, evaluating~\eqref{eq:augpost} is much faster
because it only involves diagonal and block-diagonal matrices with low-dimensional blocks.

%For the copula model of the previous section 
%for $\mY_{1:n}$, equation \eqref{eq:copmodpdf} gives the density
%\begin{align}
%  p(\yvec_{1:n};\xvec_{1:n},\varthetavec) & = c(\uvec_{1:n};\xvec_{1:n},\varthetavec) \left\{\prod_{i=1}^n \prod_{j=1}^p g_{j}(y_{i,j})\right\}.  \label{model1}
%\end{align}
%We now explain our approach to approximating the posterior for $\varthetavec$, 
%and to conducting predictive inference.

\subsection{Variational inference}
It is difficult to evaluate the complex and high-dimensional augmented posterior at~\eqref{eq:augpost} using an MCMC scheme, and we instead employ variational inference (VI) which is typically faster and more robust for such target distributions. 
Following~\citep{OngNotSmi2018}, a Gaussian approximation with a covariance matrix that has a parsimonious factor structure is used. The approximating density is
$q_\lambda(\etavec)=\phi_T(\etavec;\muvec,BB^\top+\Delta^2)$,
where $T$ denotes the length of $\etavec$, $\muvec$ is a mean vector, $B$ is
a factor loading matrix of dimension $T\times M$ where $M\ll T$ is the number of factors,
$\Delta$ is a diagonal matrix with elements $\deltavec=(\delta_1,\dots, \delta_T)^\top$. The full set of variational parameters are therefore
$\lambdavec=(\muvec^\top,\text{vec}(B)^\top,\deltavec^\top)^\top$ with $\text{vec}(B)$ denoting the vectorization
of $B$. 

At~\eqref{eq:augpost}, denote the extended likelihood term as $l(\etavec)=p(\zvec_{1:n}|\xvec_{1:n},\betavec,\varthetavec)J(\varthetavec,\xvec_{1:n})$, the prior as $p(\etavec)$, and the augmented posterior up to proportionality as $h(\etavec)=p(\etavec)l(\etavec)$.
VI proceeds by optimizing the so-called
evidence lower bound ${\cal L}(\lambdavec)$, defined as
\begin{align}
  {\cal L}(\lambdavec) & = E_q(\log h(\etavec)-\log q_\lambda(\etavec)),  \label{lowerbd}
\end{align}
where $E_q(\cdot)$ denotes the expectation with respect to $q_\lambda(\etavec)$.  Maximizing (\ref{lowerbd}) with
respect to the variational parameters $\lambdavec$ is equivalent to minimizing the Kullback-Leibler divergence
between $q_\lambda(\etavec)$ and the posterior distribution.
A common way to perform the optimization is to use stochastic gradient methods,
where we initialize with a value $\lambdavec^{(0)}$ and then update this value iteratively as
$$\lambdavec^{(k+1)}=\lambdavec^{(k)}+\avec_k\circ \widehat{\nabla_\lambdavec {\cal L}(\lambdavec^{(k)})},$$
for $k\geq 0$, where $\widehat{\nabla_\lambdavec {\cal L}(\lambdavec^{(k)})}$ is an unbiased
estimate of the gradient $\nabla_\lambdavec {\cal L}(\lambdavec)$ at $\lambdavec=\lambdavec^{(k)}$,
$\avec_k$ is a $T$-dimensional vector of step sizes at step $k$ and ``$\circ$'' denotes elementwise multiplication.
The recursion above converges to a local mode of ${\cal L}(\lambdavec)$ under conditions on the step
sizes and other regularity conditions, and updating occurs until some stopping rule is satisfied.

Among the most effective methods for computing low variance unbiased estimate of the  gradient is to use so-called reparametrization
gradients~\citep{KinWel2014,RezMohWie2014}.  
A draw from
density $q_\lambda(\etavec)$ is given by $\etavec=\muvec+B\wvec_1+\deltavec\circ \wvec_2$, with $\wvec=(\wvec_1^\top,\wvec_2^\top)^\top\sim N_{M+T}(\bm{0},I)$. Using
this generative representation \cite{OngNotSmi2018}
show the reparameterization gradient with respect to $\lambdavec$ is
\begin{align*}
  \nabla_\mu {\cal L}(\lambdavec) & = E(\nabla_\vartheta \log h(\muvec+B\wvec_1+\deltavec\circ \wvec_2)+(BB^\top+\Delta^2)^{-1}(B\wvec_1+\deltavec\circ \wvec_2)),  \\
  \nabla_{B} {\cal L}(\lambdavec) & = E(\nabla_\vartheta \log h(\muvec+B\wvec_1+\deltavec\circ \wvec_2)\wvec_1^\top+(BB^\top+\Delta^2)^{-1}(B\wvec_1+\deltavec\circ\wvec_2)\wvec_1^\top),  \\
  \nabla_{\delta} {\cal L}(\lambdavec) & = E(\text{diag}(\nabla_\vartheta \log h(\muvec+B\wvec_1+\deltavec\circ\wvec_2)\wvec_2^\top+(BB^\top+\Delta^2)^{-1}(B\wvec_1+\deltavec\circ\wvec_2)\wvec_2^\top))\,,
\end{align*}
where the expectations above are with respect to the distribution of $\wvec$.
The computations involving $(BB^\top+\Delta^2)^{-1}$ are undertaken efficiently
using  the Woodbury formula, and 
Monte Carlo estimates of the gradients are obtained using one or more Monte Carlo
samples of $\wvec$. Analytical expressions for $\nabla_\vartheta \log h(\etavec)$ are 
derived in Part~B of the Web Appendix, which we employ because it is faster and more accurate to evaluate
than automatic differentiation. 

\subsection{Predictive inference}\label{sec:predinf}
%\color{blue}
Next, we consider predictive inference for an as yet unobserved response $\bm{Y}_{n+1}$ given corresponding covariates $\xvec_{n+1}$. Let $\zvec_{n+1}=(z_{n+1,1},\dots, z_{n+1,p})^\top$, $z_{n+1,j}=\Phi^{-1}(G_j(y_{n+1,j}))$, and $S_{n+1}$ be the diagonal matrix 
defined in Section~\ref{sec:mregcop}. Then, the predictive density
at~\eqref{eq:predy} is
\begin{equation}
	p(\yvec_{n+1}|\xvec_{1:n+1},\yvec_{1:n})  =  E \left(\phi_p\left(\zvec_{n+1};S_{n+1}(I_p \otimes \xvec_{n+1})\betavec,S_{n+1}\Sigma S_{n+1}\right)\right)
	\prod_{j=1}^p\frac{g_{j}(y_{n+1,j})}{\phi_1(z_{n+1,j})}\,,\label{eq:predy2}
\end{equation}
where the expectation is with respect to $\etavec=\{\betavec,\varthetavec\}\sim p(\etavec|\yvec_{1:n})$. Given a
set of Monte Carlo draws of $\etavec$ from the variational posterior, the expectation above can be approximated in two ways.
The first is to simply average the term inside the expectation over the draws. The second is to use the draws to compute point estimates for $\etavec$ and $S_{n+1}$, and plug these into~\eqref{eq:predy2}
to replace the expectation. The latter approach 
is faster than the former, and we adopt this in our empirical work.

%We can approximate this based on a set of Monte Carlo draws $\{\betavec^{(d)},\thetavec^{(d)},\Sigma^{(d)}\}_{d=1}^D$ from the variational posterior density by computing
%$$\widehat{p}(\yvec_{n+1}\mid \xvec_{1:n+1},\yvec_{1:n})=\frac{1}{D}\sum_{d=1}^D \frac{\prod_{j=1}^p g_{j}(y_{n+1,j})}{\prod_{j=1}^p \phi(z_{n+1,j})}
%\phi_p\left(\zvec_{n+1};S_{n+1}^{(d)}(I_p \otimes \xvec_{n+1})\betavec^{(d)},S_{n+1}^{(d)}\Sigma^{(d)} S_{n+1}^{(d)}\right),$$
%where
%\begin{align*}
%S_{n+1}^{(d)}=\mbox{diag}\left((1+\xvec_{n+1}P_1(\thetavec_1^{(d)})^{-1}\xvec_{n+1}^\top)^{-1/2},\dots,(1+\xvec_{n+1}P_p(\thetavec_p^{(d)})^{-1}\xvec_{n+1}^{\top})^{-1/2}\right).
%\end{align*}
%
%{Alternatively, we can approximate the posterior predictive density based on point estimates for $\betavec$, $\thetavec$ and $\Sigma$.  Denote the corresponding variational
%posterior mean values of these parameters by $\overline{\betavec},\overline{\thetavec},\overline{\Sigma}$.  Plugging in these estimates 
%gives the posterior predictive density estimate 
%$$\widehat{p}_{\text{plug-in}}(\yvec_{n+1}\mid \xvec_{1:n+1},\yvec_{1:n})=\frac{\prod_{j=1}^p g_{j}(y_{n+1,j})}{\prod_{j=1}^p \phi(z_{n+1,j})}
%\phi_p\left(\zvec_{n+1};\overline{S}_{n+1} (I_p \otimes \xvec_{n+1})\overline{\betavec},\overline{S}_{n+1}\overline{\Sigma}\,\overline{S}_{n+1}\right),$$
%where
%\begin{align*}
%\overline{S}_{n+1}=\mbox{diag}\left((1+\xvec_{n+1}P_1(\overline{\thetavec}_1)^{-1}\xvec_{n+1}^\top)^{-1/2},\dots,(1+\xvec_{n+1}P_p(\overline{\thetavec}_p)^{-1}\xvec_{n+1}^\top)^{-1/2}\right).
%\end{align*}
%}

\subsection{Marginal regression functions}
The $p$ marginal regression functions for the pseudo-response and response
variables are defined as $m_j(\xvec_{n+1})=E(Z_{n+1,j}|\xvec_{1:n+1},\yvec_{1:n})$ and $f_j(\xvec_{n+1})=E(Y_{n+1,j}|\xvec_{1:n+1},\yvec_{1:n})$, respectively, for $j=1,\ldots,p$. These can be expressed as 
\[
m_j(\xvec_{n+1})=\int s_{n+1,j} \xvec_{n+1}\betavec_j p(\etavec|\yvec_{1:n})d\etavec\,, \mbox{ and }
\]
\begin{equation*}\begin{aligned}
E(Y_{n+1,j}|\xvec_{1:n+1},\yvec_{1:n}) =\int\int G_{j}^{-1}(\Phi(z_{n+1,j}))\phi_1(z_{n+1,j};s_{n+1,j} \xvec_{n+1,j}\betavec_j,(s_{n+1,j}^2))d z_{n+1,j}\, p(\etavec|\yvec_{1:n})d\etavec,
\end{aligned}\end{equation*}
where
$s_{n+1,j}=(1+\xvec_{n+1}P_j(\thetavec_j)^{-1}\xvec_{n+1}^\top)^{1/2}.$
Estimates of these marginal mean functions can be obtained by approximating
the expectations in the above integrals by Monte Carlo samples from 
the variational posterior, or by plugging in point estimates, 
similar to the approximation of the predictive densities.

\section{Australian Electricity Prices} \label{sec:electricity}
%We apply our methodology to high-frequency electricity prices in the Australian National Electricity Market (NEM).
There is an extensive 
literature concerned with modeling and forecasting high-frequency intraday electricity prices in wholesale electricity markets~\citep{weron2014}.
Recent research includes understanding how economic fundamentals affect
these; for example, see \cite{smithshively2018} and \cite{yantruck20}. However, complicating the problem is that 
in many markets prices vary regionally. Our proposed multivariate distributional regression approach provides an effective means to account for this.
\subsection{Problem description}
We apply our multivariate response regression copula model to half-hourly prices in the 
inter-connected regions of the Australian National Electricity Market (NEM). These regions coincide with
the 
states of New South Wales (NSW), Queensland (QLD), South Australia (SA), Tasmania (TAS) and Victoria (VIC). 
The advantage of using our model for this problem
is that it can account for three known key features in the data: (i) the complex non-Gaussian marginal
 distributions of regional prices, 
(ii) the nonlinear and multivariate relationship of regional demands with each individual regional price, and (iii) the strong inter-regional  dependence in prices due to unobserved supply side factors.

\subsection{Data and model}
The data are the 17,250 half-hourly electricity prices and demand observed in each of the $p=5$ regions during 2019. 
In the NEM, prices can be negative with 
a floor price of $-\$1000$, so we follow~\cite{smithshively2018} and~\cite{yantruck20} and set $Y_{i,j}=\ln(1001+\mbox{Price}_{i,j})$,
where $\mbox{Price}_{i,j}$ is the price in region $j$ at half-hour $i$. 
The margins $G_1,\ldots,G_5$,
are estimated using adaptive kernel
density estimators bounded to the feasible region for transformed prices, which are highly non-Gaussian; 
see Part~C of the Web Appendix for these estimates. 
The design matrix $F$ at~\eqref{eq:linmod} is constructed using a multivariate basis expansion of the five regional demand variables.
In particular, a cubic thin plate spline basis~\citep{Woo2003} is used with $k=30$ knots set equal to the centroids of demand clusters
obtained from a
$k$-means clustering algorithm.\footnote{This is a popular way to select multivariate  knots for such a radial basis because it ensures the basis design
follows the data distribution of the covariates.}
There are a total of $q=50$ basis functions (so that $\xvec_i$ has 50 elements), where the first 20 basis functions are the polynomials of degree less than three.
\begin{figure}[thbp]
	\centering\includegraphics[width=0.5\textwidth]{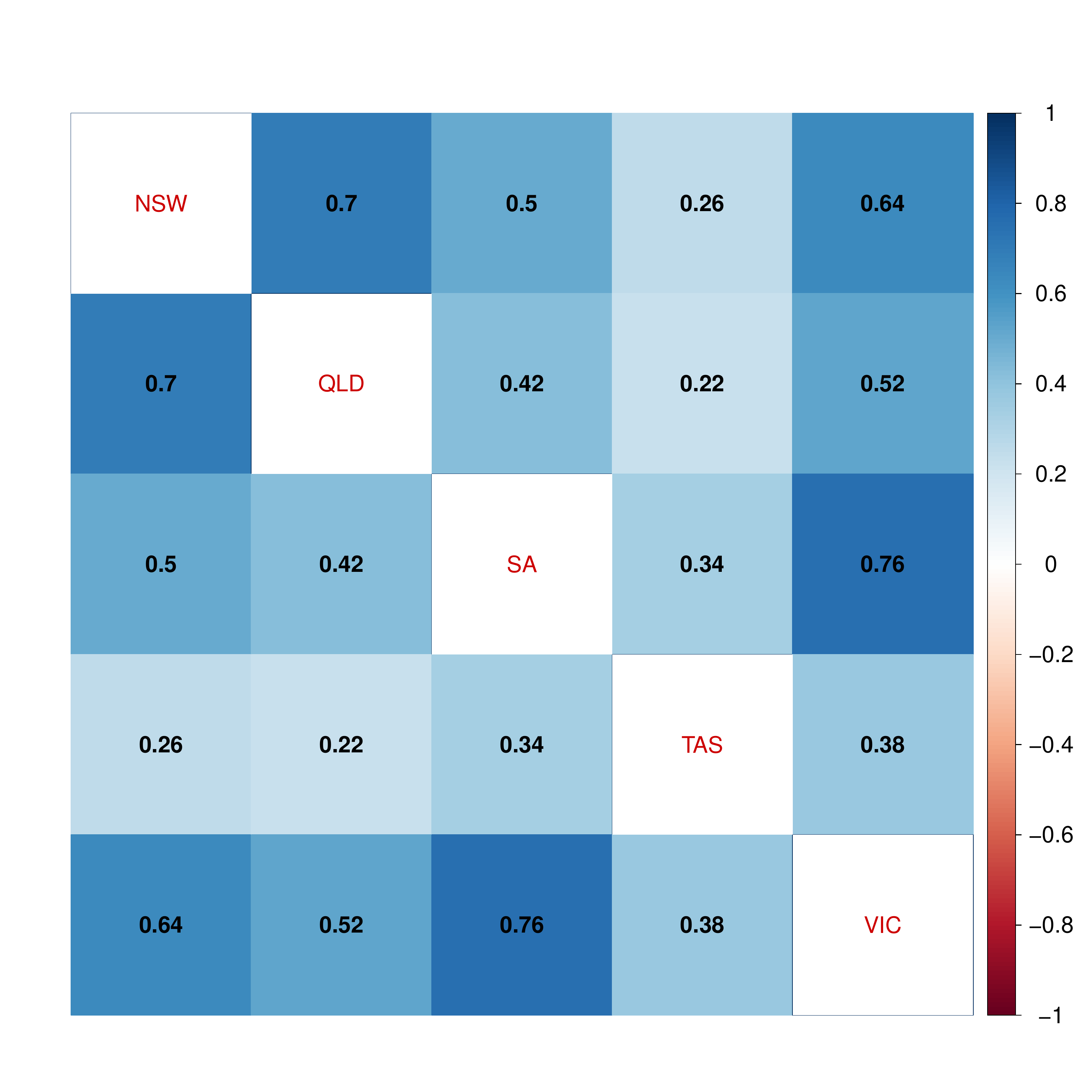}
	\caption{\footnotesize \emph{Electricity prices.} Estimated inter-regional Spearman correlation matrix $\Gamma^S$ for our regression copula model fit to the
		half-hourly Australian NEM electricity price and demand data from 2019. The matrix $\Gamma^S$ is evaluated at the
		median values of regional demand and using variational estimates of the  copula model parameters.} 
	\label{fig:VSpearman}
\end{figure}

\subsection{Empirical results}
\paragraph{Inter-regional dependence} We use the regression copula at~\eqref{eq:gcop}, evaluated at the (variational) posterior mean of 
$\varthetavec=\{\Sigma,\thetavec\}$, to measure the 
inter-regional dependence in prices. Because it is a Gaussian copula process (i.e.~it is a function of 
the covariates), we evaluate it at the covariate value $\xvec_{\mbox{\tiny med}}$ with elements computed at the median value of demand in each region.
Figure~\ref{fig:VSpearman} plots the Spearman correlation matrix 
$\Gamma^S = \frac{6}{\pi}\arcsin(R_{\mbox{\tiny med}}/2)$, 
%$\Gamma^S = \frac{6}{\pi}\arcsin(R_{\mbox{\tiny med}}/2)$.
where $R_{\mbox{\tiny med}}=S_{\mbox{\tiny med}}V_{\mbox{\tiny med}}S_{\mbox{\tiny med}}$, 
$V_{\mbox{\tiny med}}=\Sigma + (I_p \otimes \xvec_{\mbox{\tiny med}}) (\Sigma \star P(\thetavec)^{-1})(I_p \otimes \xvec_{\mbox{\tiny med}}^\top)$
and $S_{\mbox{\tiny med}}=\mbox{diag}(V_{\mbox{\tiny med}})^{-1/2}$ are the similarly denoted terms in Section~\ref{sec:copprocess} evaluated
at $\xvec_{\mbox{\tiny med}}$. 
The prices are positively dependent, with the level quantifying 
the degree of market integration. For example, the  pairs (NSW, QLD), (NSW, VIC) and (VIC, SA)  have the highest Spearman correlations, which is
consistent with these being geographically adjacent regions connected by  high voltage direct current inter-connectors. In contrast, TAS has the lowest Spearman correlations with the other states, which is because this island is the least integrated with the rest  of the NEM
due to very different supply 
side factors (e.g.~weather and generator mix) and limited inter-connector capacity with the mainland. 

\begin{figure}[htbp]
	\centering\includegraphics[width=0.8\textwidth]{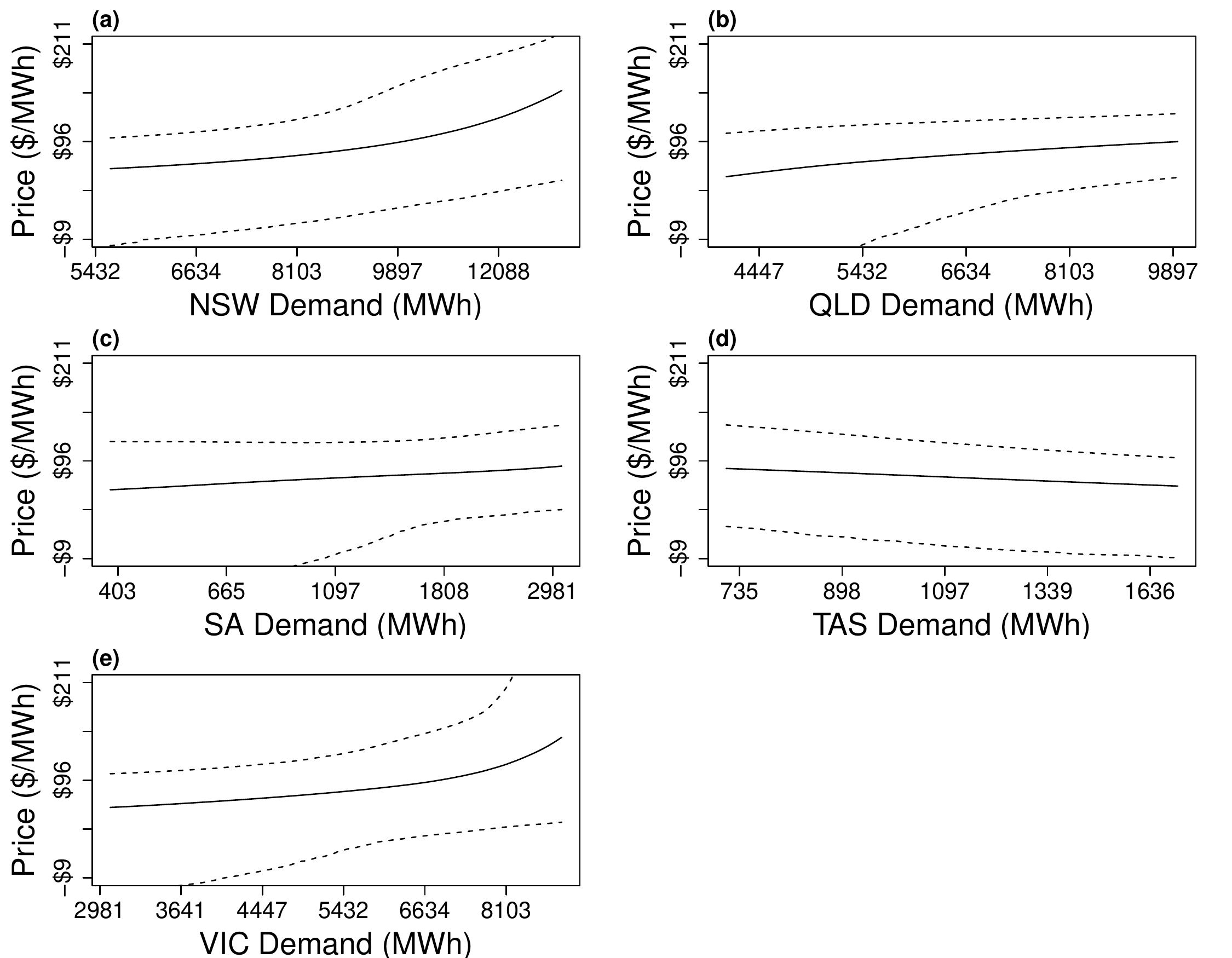}
	\caption{\footnotesize  \emph{Electricity prices.} The solid line gives the expected value of the predictive distribution of system-wide price $\bar Y$  
		given regional demand values. Each panel contains the a slice of this multivariate function obtained by varying demand in one region, while holding demand in the other regions fixed to their median values. The vertical axis is given in \$ per megawatt hours,
		and dashed lines give 90\% credible intervals.}
	\label{fig:IRelect}
\end{figure}
\paragraph{Effect of demand on price} System-wide price is measured using the demand-weighted\footnote{The weights
	are the normalized reciprocal of total annual demand in each region, given by 0.3687 (NSW), 0.2355 (VIC), 0.2818 (QLD),
	0.0624 (SA) and 0.0516 (TAS). 
}  average of regional (logarithmic) prices, $\bar Y$.
Using this measure, we compute a multivariate regression function as the
 expected value of the predictive distribution of $\bar Y$, 
given regional demand values. This is done via sampling from the joint plug-in posterior predictive density and then computing $\bar Y$. 
%Note that a multivariate regression model is
%necessary to do so, because it allows for the strong dependence between regional prices.
Figure~\ref{fig:IRelect} plots ``slices'' of this function obtained by varying demand in one region, while holding demand in the 
other regions fixed to their median values. 
For example, panel~(a) gives the system-wide impact of demand variation in NSW, holding the
demand values in the other four regions constant. Demand in the largest regions (NSW, QLD and VIC) has the greatest impact, 
and in NSW and VIC the relationship is nonlinear. 
This is because when demand exceeds baseline supply, much more expensive short run ``peaking'' generation capacity 
is required to meet demand.
% \citep{smithshively2018}.

\begin{table}[htbp]
	\renewcommand{\arraystretch}{1.15}
	\begin{center}
		\caption{\footnotesize \emph{Electricity prices.} Electricity demand at midday on 15 Jan and 15 July 2019} 
		\begin{tabular}{lcccccc}
			\hline\hline
			&\multicolumn{6}{c}{Region} \\ \cline{2-7} 
			Date &NSW &QLD &SA &TAS &VIC &NEM \\ \hline
			15 Jan 19 &11,068	&6,955	&1,984	&1,264	&7,697	&28,969\\
			15 July 19 &8,466	&5,317	&1,515	&1,282	&5,621	&22,201\\ \hline\hline
		\end{tabular}
		\label{tab:demand}
	\end{center}
	\centering{\footnotesize\em Note: demand is reported in MWh for each region, along with total system demand.}
\end{table}

\begin{figure}[htb]
	\centering\includegraphics[width=0.8\textwidth]{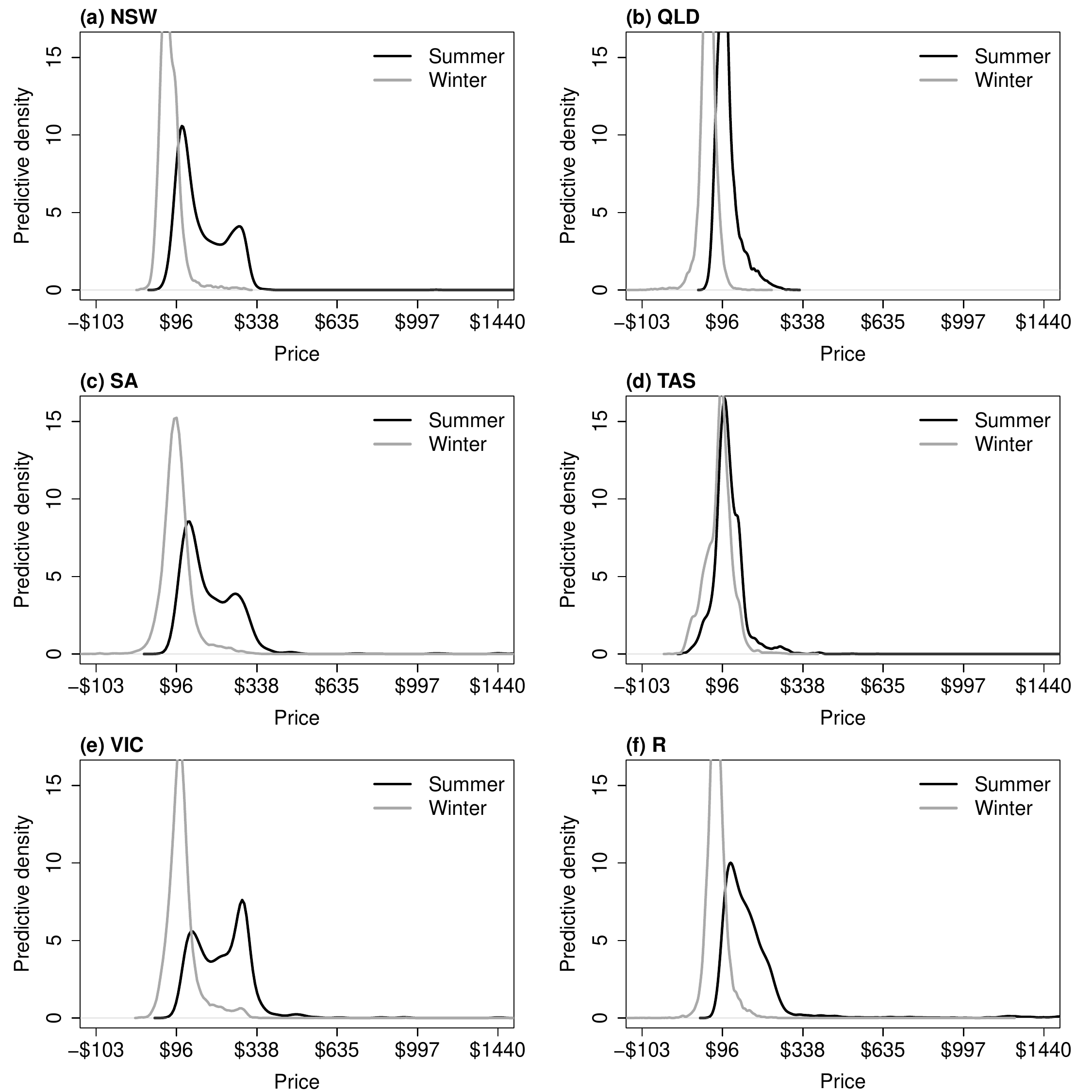}
	\caption{\footnotesize Predictive densities for electricity prices at midday on 15 Jan 2019 (labelled `summer') and 15 July 2019 (labelled `winter'). Panels~(a) to (e) give the densities for prices in each of the five regions, and panel~(f) for the demand-weighted system price. The horizontal axis is on the logarithmic scale.} 
	\label{fig:electpred}
\end{figure}

\paragraph{Predictive distributions} To illustrate the flexibility of the predictive distributions, they are computed for regional prices and system-wide price $\bar Y$ at two time points:
midday on 15 January (mid-summer) and 15 July 2019 (mid-winter). Table~\ref{tab:demand} reports electricity demand at these two times, which is higher in every region (except TAS) on 15 January due to 
climatic conditions.\footnote{It was particularly hot on 15 January across the Australian mainland (but not TAS) resulting in very high air-conditioning load.}  
Figure~\ref{fig:electpred} plots the predictive distributions of all regional prices and $\bar Y$, and
those on 15 January have higher means and spread, along with 
extenuated upper tails. This is consistent with a known feature of the NEM, where during 
periods of high demand there is a sizable
increase in upper tail risk in electricity prices.

\subsection{Benchmarking}
To validate our model we compare its predictive accuracy for $\bar Y$ 
with those from other multivariate regression models using 10-fold cross-validation. 
The root mean square error (RMSE) is used to measure point prediction accuracy, and
the log-score (LS) and cumulative ranked probability score (CRPS) measure distributional predictive accuracy. The competing models are:
\begin{itemize}
	\item MVC.prior1 and MVC.prior2: our proposed approach using priors~1 and~2 for $\Sigma$.
	\item MVC.add.prior and MVC.add.prior2: as above, but with an additive basis for demand covariates, with $F$ comprising univariate thin plate spline basis terms.
	\item MVC.lin.prior1 and MVC.lin.prior2: as above, but with a linear basis over demand covariates, so that $F$ comprises only linear terms.
	\item MVN.lin: a Gaussian SUR model with linear regression terms.
	\item MVN.lin.het and MVN.lin.cov: conditionally Gaussian SUR model with linear regression terms and the 
	marginal variances (lin.het) and also entries of a Cholesky factor of the correlation matrix (lin.cov) being
	linear functions of the demand covariates.
	\item MVN.add, MVN.add.het and MVN.add.cov: same as the MVN models, but with additive functions of the demand covariates using univariate thin plate splines.
	% instead of linear functions.
	\item NOC: A Gaussian copula model with the marginals for each regional price given by multivariate regressions using the same basis as MVC.prior1/MVC.prior2. 
\end{itemize}
The models denoted MVC are the proposed regression copula models,
but with differing function bases. 
The models denoted MVN are the conditionally
Gaussian  models of \citet{MusMaySimUmlZei2022} estimated using the R-package ``\texttt{mvnchol}''. 
The model denoted NOC employs the covariates to define nonlinear regressions in the marginals, as in~\cite{PitChaKoh2006,song2009}, rather
than to specify the copula process.

 \begin{table}[tbh]
 	\renewcommand{\arraystretch}{1}
 	\begin{center}
 		\caption{\footnotesize \emph{Electricity prices.} Australian NEM system-wide electricity price predictive accuracy}
 	\begin{tabular}{lccc}
 		\hline \hline
 		& CRPS & LS & RMSE \\ 
 		\hline
 		\multicolumn{4}{l}{Multivariate Regression Copula Models}\\ \hline
 		MVC.prior1 & \textbf{0.01560} & \textbf{-2.23568} & \textbf{0.00167} \\ 
 		MVC.prior2 & 0.01563 & -2.24026 & \textbf{0.00167} \\ 
 		MVC.add.prior1 & 0.01560 & -2.23422 & 0.00178 \\ 
 		MVC.add.prior2 & 0.01568 & -2.23519 & 0.00178 \\ 
 		MVC.lin.prior1 & 0.01856 & -1.98366 & 0.00228 \\ 
 		MVC.lin.prior2 & 0.01849 & -2.00912 & 0.00226 \\ 
 		\hline
 		\multicolumn{4}{l}{Conditionally Gaussian Models}\\ \hline
 		MVN.lin & 0.01809 & -1.45180 & 0.00182 \\ 
 		MVN.lin.het & 1.29217 & -0.04494 & 0.50086 \\ 
 		MVN.lin.cov & 1.95226 & 0.21337 & 36.71676 \\ 
 		MVN.add & 0.16709 & 0.58704 & 0.00161 \\ 
 		MVN.add.het & 0.45309 & -0.43186 & 0.12253 \\ 
 		MVN.add.cov & 0.39934 & -0.14563 & 0.29181 \\ 
 		\hline
 		\multicolumn{4}{l}{Gaussian Copula with Regression Marginals}\\ \hline
 		NOC & 0.01807 & -2.01589 & 0.00199 \\ 
 %		NOC.dagum & 0.15300 & -0.60805 & 0.24100 \\ 
 		\hline \hline
 	\end{tabular}
 \label{tab:EPD:scores}
\end{center}
\centering{\footnotesize \em Rows give results using different distributional regression methods. Predictive accuracy is computed using
 10-fold CV, and measured using RMSE (point), LS and CRPS (distributional). Lower values of all metrics
 correspond to higher accuracy. The approach proposed in this paper is most accurate.}
 \end{table}
 
 Table~\ref{tab:EPD:scores} reports the predictive accuracy metrics and we observe that (i)~Prior~1 outperforms Prior~2, (ii)~our proposed MVC models dominate, (iii) the MVC models that capture the relationship between demand and price as nonlinear and multivariate are most accurate, and (iv) it is better to capture the impact of demand through the regression copula, rather than through
 regression marginals as in the NOC model. 

\section{Likelihood-free Inference for Tree Species Abundance}\label{sec:ecology}

We now consider an application of our methodology to likelihood-free
inference (LFI) for  
tree species abundance survey data.
The data are from \citet{Chretal2014}, and are species counts 
from five complete censuses of trees for a period spanning 1987-2005
in a 50 ha plot in Pasoh, Malaysia.  Data from a second site 
is used for setting an informative prior.

We first describe the stochastic model considered by \cite{Chretal2014}.   
This is followed by a discussion of LFI and
how to use distributional regression to perform such inference.  
Finally, we present the results of our analysis.  
We compare these to univariate marginal posterior estimation of 
\cite{KleNotSmi2021}, and these authors benchmark
their approach against a range of alternative LFI 
methods.  The focus is on how multivariate estimation improves
inference in this example, compared to univariate estimation.

\subsection{Model for tree species abundance}
Consider a forested area (a “site”) that is censused at various intervals to track the abundances of the tree species present.
Let $N_{it}$ denote the abundance of tree species $i$ for census $t$, for
$t=1,\dots, T=5$ and
$i=1,\dots,I=814$.  There are $T-1$ census intervals with durations
$\Delta T_{it}$, for 
$t=1,\dots, T-1$, $i=1,\dots, I$.  The census intervals are species specific, 
since the average census time for trees of a given species depends on
their spatial distribution.  
Let $S_{it}$ denote the number of trees of species $i$ surviving 
within census interval $t$.
Define $A_{it}=N_{i(t+1)}-S_{it}$, so that $A_{it}$ represents recruitment in census interval $t$.
\citet{Chretal2014} model the data for each census interval separately, 
conditionally on the initial abundance for the interval.    
For census interval $t=1,\ldots,T-1$, 
the model (conditional on $N_{it}$, $i=1,\dots, I$) is
\begin{align}
  S_{it}|N_{it} & \sim \text{Binomial}(N_{it},\exp(-\mu_{it}\Delta T_{it})), \label{survival} \\
  A_{it}|S_{it},N_{it} & \sim \text{Poisson}\left(N_{it}\left\{ \exp(\rho_{it}\Delta T_{it})-\exp(-\mu_{it}\Delta T_{it})\right\}\right),
  \label{recruitment}
\end{align}
where the parameters $\mu_{it},\rho_{it}$
are instantaneous mortality and growth rates, respectively. The mortality and growth rates are given hierarchical priors,
\begin{align*}
  \mu_{it}|\rho_{it}  & \sim \text{LN}(\Omega_{t},\sigma_{t}^2)I(\mu_{it}>-\rho_{it}), \;\;\;\;
  \rho_{it} \sim \text{ALD}(c_{t},\phi_{1t},\phi_{2t}),
\end{align*} 
where $\text{LN}(\mu,\sigma^2)I(A)$ denotes a log-normal distribution with parameters $\mu$ and $\sigma^2$, truncated to region $A$, and
$\text{ALD}(c,\phi_1,\phi_2)$ denotes an asymmetric Laplace distribution with density
\begin{align*}
  f(x;c,\phi_1,\phi_2) & = \left\{ \begin{array}{cc}
    k\exp((x-c)\phi_2) & \mbox{$x<c$} \\
    k\exp((c-x)\phi_1) & \mbox{$x\geq c$},
    \end{array}\right.
\end{align*}
and $k=\phi_1\phi_2/(\phi_1+\phi_2)$.  The prior for $\rho_{it}$ is truncated 
to ensure
that the mean of the Poisson distribution in \eqref{recruitment} is positive. 
The priors on the hyperparameters $\Omega_{t}$, $\sigma_{t}^2$, $c_{t}$, 
$\phi_{1t}$ and $\phi_{2t}$ are described later.  
\citet{Chretal2014} use MCMC to evaluate the posterior of the parameters for each census interval, where for computational tractability 
an approximation of the likelihood was used.

\subsection{Likelihood-free inference}
LFI is used to compute Bayesian inference for models where
computing the likelihood is  impractical.  Suppose we have
data $\mathcal{D}$ with observed value denoted $\mathcal{D}_{\text{obs}}$, and
a model for the data involving parameters $\psivec$, specified by a density
$p({\mathcal{D}}|\psivec)$.  The prior density for $\psivec$ is $p(\psivec)$.  
For the joint density 
$p(\psivec,{\mathcal{D}})=p(\psivec)p({\mathcal{D}}|\psivec)$, the
conditional density for $\psivec$ given 
${\mathcal{D}}={\mathcal{D}}_{\text{obs}}$ is the
posterior density $p(\psivec|{\mathcal{D}}_{\text{obs}})$.  From this observation, 
if we simulate samples $\{(\psivec_i,{\mathcal{D}}_i)\}_{i=1}^n$ independently
from $p(\psivec)p({\mathcal{D}}|\psivec)$, a distributional
regression model can be fitted using these simulations as training data to estimate
$p(\psivec|{\mathcal{D}}_{\text{obs}})$.  When using the distributional
regression model outlined here, 
set $\bm{Y}_i=\psivec_i$, $\xvec_i={\mathcal{D}}_i$, and consider the multivariate predictive density
of $\bm{Y}_{n+1}$ given $\xvec_{n+1}={\mathcal{D}}_{\text{obs}}$ as the estimate
of the multivariate posterior density. 
This procedure only requires the simulation of the training data from
the model, and not evaluation of the likelihood $p({\mathcal{D}}|\psivec)$.
  
Often LFI methods do not use the raw data ${\mathcal{D}}$ for estimating the 
posterior density, but instead use a lower-dimensional 
summary statistic vector $\mH_{\text{obs}}=\mH({\mathcal{D}}_{\text{obs}})$. 
There can be two advantages to doing this.  The first is computational, where
the reduction in dimension (without losing too much information about
$\psivec$) can also reduce Monte Carlo variation of posterior estimates
using LFI algorithms.  The second is to make 
inference robust to the misspecification of the stochastic model. 
\cite{Lewis2021} recently highlight and greatly develop the ``restricted likelihood''
approach to Bayesian inference with misspecified models, where
a posterior distribution is constructed by conditioning on 
a summary statistic rather than the full data.  
The use of a summary statistic allows us to discard information that cannot
be matched under the assumed model, with the intended use of the model
able to guide what summaries the analyst should focus on matching.  
Even in models with tractable likelihood, the likelihood for summary statistics
may be intractable, so that LFI methods are attractive for restricted
likelihood computation~\citep{nott+df24}.
 
We implement a restricted likelihood approach for analyzing
the tree species abundance model, where there
is concern about the adequacy of the binomial and Poisson models for
survival and recruitment.  
We consider LFI based on simulation of 
data replicates  from (\ref{survival}) and (\ref{recruitment}), after simulating parameters from the prior.  We use 
the following hyperpriors:  
\begin{equation*}\begin{aligned}
\Omega_{t}&\sim N(-3.3,1.2^2), &c_{t}&\sim N(0.07,0.08^2),\\
\sigma_{t}^2&\sim IG(2,1),
&\phi_{1t}, \phi_{2t} &\sim IG(3,200),
\end{aligned}\end{equation*}
$t=1,\dots T-1$, where $IG(a,b)$ is an inverse gamma distribution with shape $a$ and scale $b$.  The hyperpriors
were chosen after
examining estimates of growth and mortality rates from
another site (Barro Colorado Island in Panama), devising priors
summarizing the variation of these estimates, and then making 
these more dispersed to account for inter-site differences 
and to avoid any resulting prior-data conflicts.  The data used 
for both sites can be found in the supplementary materials of \citet{Conetal2006}.

We use our multivariate regression model to estimate the joint
posterior distribution of
the parameters $\Omega_{t},\sigma_{t}^2,\phi_{1t},\phi_{2t}$, $t=1,\dots, T-1$.
These are the parameters
of main scientific interest in the model, as they enable a crude estimate of the amount of variation
in abundance fluctuations due to environmental variance and to demographic variance respectively
at different times.
Environmental variance is variability due to temperature, rainfall, pests and other environmental factors.
Demographic variance is variability due to the stochastic nature of birth and death processes and
variation of birth and death rates within a population due to individual specific factors.  There are
other relevant sources of variation also,
see \citet{Chretal2014} for
further discussion.

Temporal environmental variability is correlated across individuals, and if this is the main
factor driving abundance fluctuations it is expected that the variance of abundance fluctuations
should scale roughly quadratically with initial abundance.  On the other hand, if demographic variability
drives abundance levels, simple population dynamics models exhibit variance of abundance changes
scaling linearly with abundance levels.  In practice, a scaling exponent between $1$ and $2$ is
usually observed and for the data considered here scaling exponents near $2$ provide a good fit
for common species, while values near $1$ provide a good fit for rare species.  This suggests that
environmental variability is driving abundance levels for common species while demographic variability
is most important for rare species, consistent with theoretical expectations.

In discussing limitations of their model, \citet[p.~5]{Chretal2014} mention that the assumption
of binomial and Poisson distributions for survival and recruitment may be too simple.
Hence our motivation for using LFI here is to fit directly to scientifically meaningful summary
statistics so that the fitted model is fit-for-purpose for understanding sources
of variability driving population dynamics.  The summary statistic likelihood
is intractable, so LFI methods are essential for computation.  
The summary statistics we define relate to direct estimates of 
scaling exponents of
variability of abundance fluctuations with initial abundance, as well as
overall growth and mortality.  
A detailed description of the summary statistics is given in the supplementary
material.  We will see that a restricted likelihood analysis
 allows the model to match the summaries, whereas a conventional
 Bayesian analysis based on the full data using the approximate 
 full likelihood method in \citet{Chretal2014} does not.

\subsection{Distributional regression for posterior approximation}
We consider analyses to approximate the joint posterior density for
$\psivec_{t}=(\Omega_{t},\sigma_{t}^2,c_{t},\phi_{1t},\phi_{2t})^\top$ using summary statistics $\mH_{t}$ for census interval $t=1,\ldots,T-1$. In forming the regression predictor in \eqref{eq:linmod} we use  a thin plate regression spline basis expansion as in Section \ref{sec:electricity} with $k=50$ knots chosen as the centroids of a $k$-means 
clustering. With a smoothness penalty of order 3 we thus arrive at 70 basis functions.

We are interested in inference on $\psivec_{t}$
because it gives an approximate but interpretable
partitioning of the abundance variation into environmental and demographic components at time $t$, leading to
\begin{equation}\label{eq:var}
\mbox{Var}(N_{t+1}\mid N_{t}=N)\approx N^2 \Delta T_{t}^2 v_e+N\Delta T_{t} v_d
\end{equation}
where $v_e=\mbox{Var}(\rho_{t})$ and $v_d=E(\rho_{t})+2E(\mu_{t})$, see (S8) in the supporting materials of
\citet{Chretal2014}.  Following these authors we estimate $\psivec_{t}$ at each census interval separately and compare the original approximate likelihood-based estimates of \citet{Chretal2014} (denoted by CTFS, 
obtained using the CTFS R package) with our multivariate regression copula with priors 1 and 2 (denoted by MVC prior1, MVC prior 2) and the univariate regression copula approach of \citet{KleNotSmi2021} (denoted by UVC). The latter three LFI methods use the summary statistics and are trained with $n=5,000$ samples   $\lbrace\psivec_{t}^{(r)},{\mathcal{D}}_{t}^{(r)}\rbrace_{r=1}^n$ of parameters and data sets ${\mathcal{D}}_{t}^{(r)}=\lbrace S_{1t}^{(r)},A_{1t}^{(r)},\ldots,S_{It}^{(r)},A_{It}^{(r)}\rbrace$ of the same size as the observed data.
We train MVC, UVC with 30,000 VI iterations, $K=1$ and $M=10$. Unlike many other LFI methods, our approach is ``amortized", which means that after model training, inference and prediction on any new data set can be performed directly.  For evaluation of the results, we generated 1,000 additional test samples. 

\subsection{Results}
\begin{sidewaysfigure}[htbp]
	\centering\includegraphics[width=0.9\textwidth]{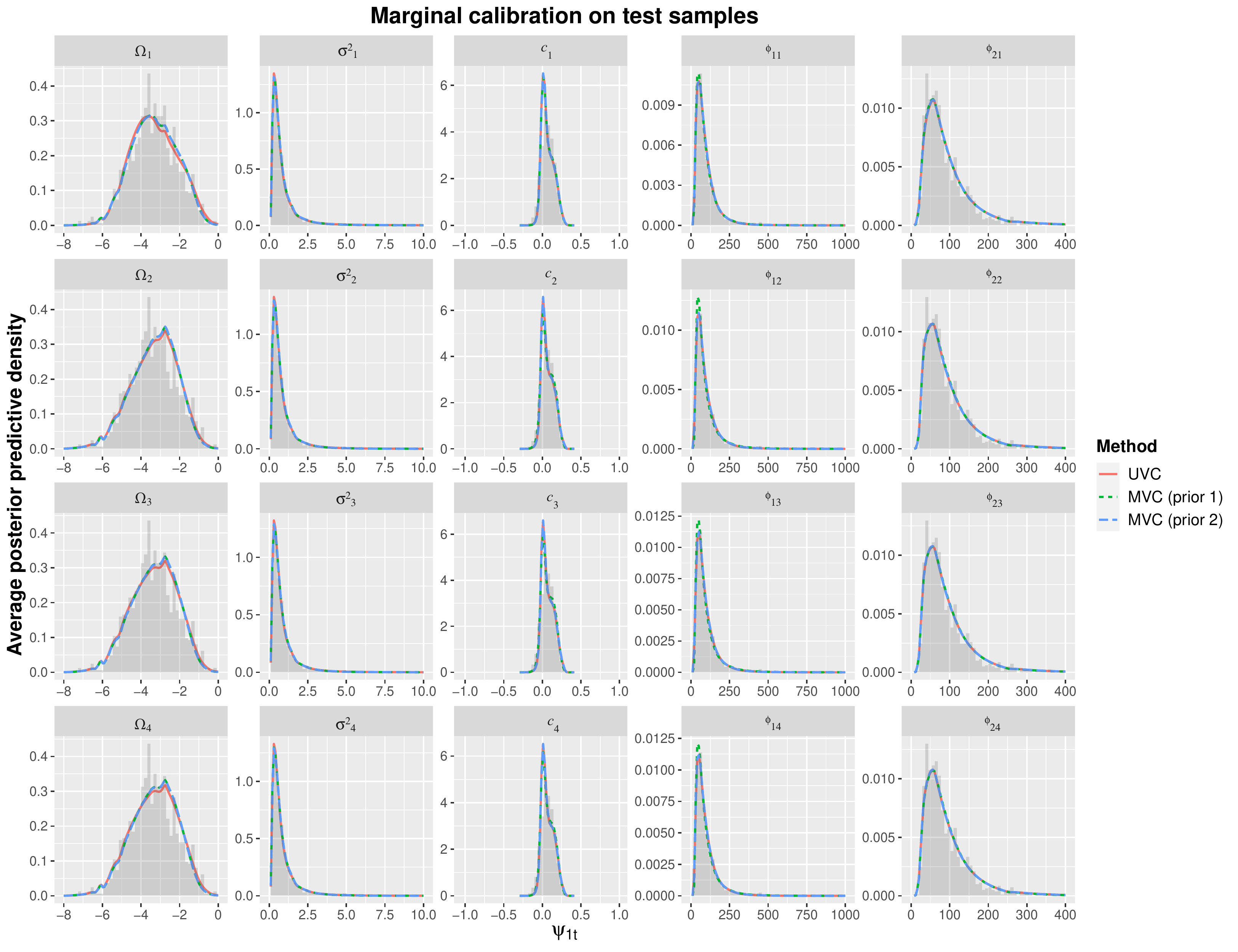}
	\caption{\footnotesize \emph{Ecology data.} Plots of average estimated 
		posterior densities for UVC and MVC with priors 1 and 2 (solid lines) 
		and prior samples (grey histograms).} 
	\label{fig:ecology:margcal}
\end{sidewaysfigure}

\paragraph{Marginal calibration} 
One criterion for evaluating the probabilistic forecast at~\eqref{eq:predy2}  is its ``calibration", which
is its statistical consistency with the observations~\citep{Gneetal2007}.  
\cite{KleNotSmi2021} discuss
marginal calibration properties of probabilistic
forecasts from distributional regression methods, and demonstrate
that the univariate version of our proposed copula
method has good marginal calibration properties.
They also discuss the application of regression copula distributional
regression for LFI, although they are only able
to produce estimates of univariate marginal posterior distributions because
their method is univariate.  For LFI, 
marginal calibration 
corresponds to 
the average estimated posterior distribution, for datasets drawn from the prior
predictive distribution, being equal to the prior.  This calibration property holds
for the true posterior density, since we can write
$$p(\psivec)=\int p(\psivec)p({\mathcal{D}}|\psivec)\,d{\mathcal{D}} = \int p({\mathcal{D}})p(\psivec|{\mathcal{D}})\,d{\mathcal{D}},$$
showing that the expectation of the true posterior density 
is the prior, averaging over data drawn from the prior predictive.  
For an LFI method, 
although good calibration doesn't guarantee accuracy it is certainly
desirable, and poor calibration indicates possible posterior inaccuracy.  
Figure \ref{fig:ecology:margcal} demonstrates that both the UVC method
and also our proposed
MVC method are marginally calibrated.
It plots
the average posterior density estimates (solid lines) from the posterior predictive densities in the test samples versus a probability
histogram of the prior.  

\begin{figure}[htb]
\centering\includegraphics[width=0.9\textwidth]{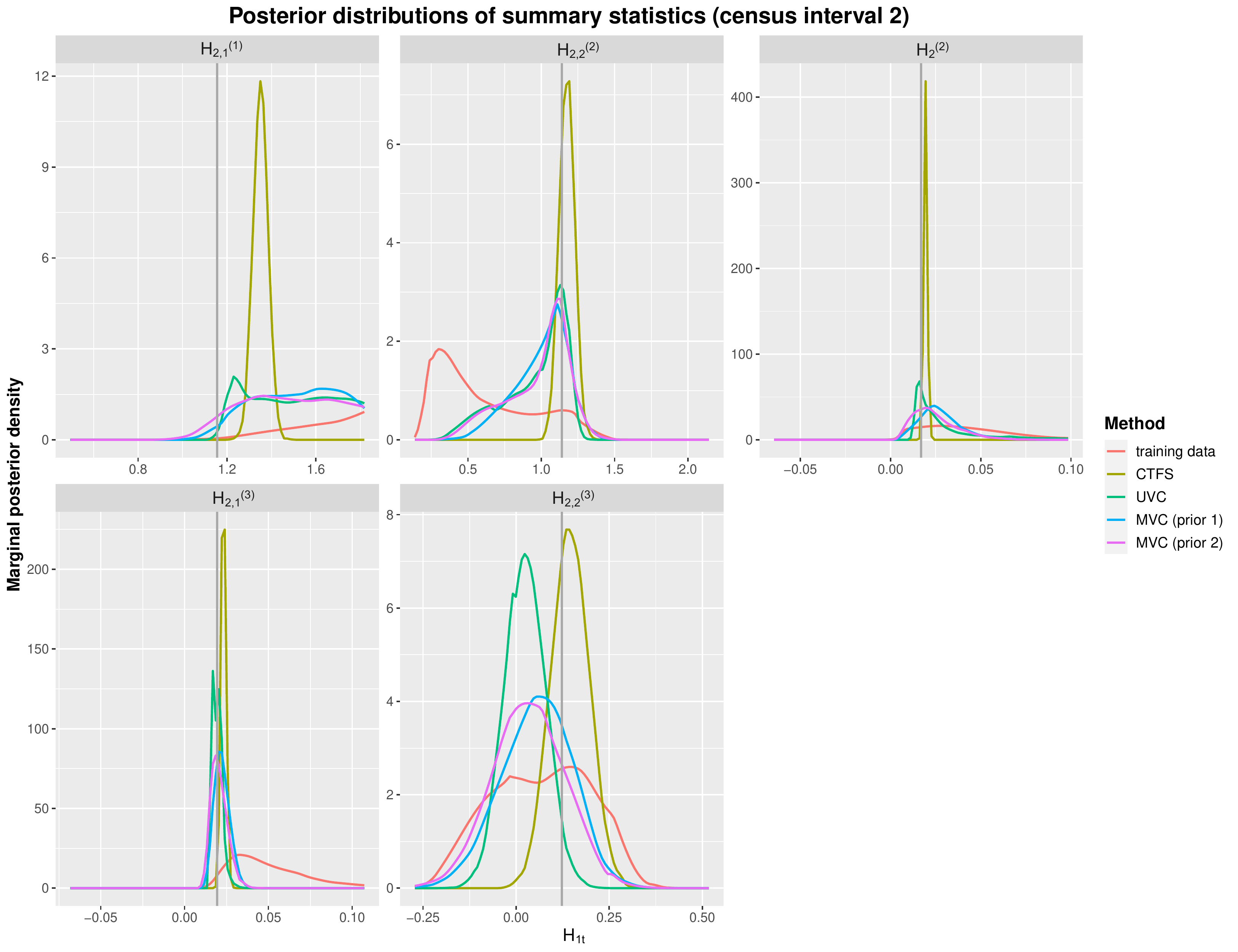}
\caption{\footnotesize \emph{Ecology data.} Posterior predictive densities of five summary statistics for census interval 2.  The first summary statistic considered
(top left graph) is an estimate of the scaling exponent of variance of abundance
fluctuations with initial abundance.} 
\label{fig:ecology:post:summ}
\end{figure}

\paragraph{Posterior estimation and posterior predictive densities of summaries}
Figure B in the supplementary material 
shows the estimated marginal posteriors 
 together with a histogram of samples from the priors for all census intervals. 
 Generally, the MVC and UVC methods agree
 well with each other, but the estimates from the CTFS method without summary statistics 
differ.  
 Figure \ref{fig:ecology:post:summ} summarizes the posterior
predictive densities of summary statistics obtained by 10,000 samples 
from the estimated posterior distributions for each method and for census 
interval 2.  The vertical lines indicate the observed
summary statistic values.  For this census interval, 
it can be seen that the likelihood-based full data method 
CTFS produces a posterior predictive density for the first summary statistic
for which the observed value is out in the tails, indicating that the fitted model
cannot match the observed value.  A similar result can be seen for
census interval 4 (results not shown).  The first summary statistic
is a pooled estimate of the scaling exponent of variance abundance fluctuations with initial abundance, 
which is an important quantity in the scientific study.  
It is expected on theoretical grounds that the scaling exponent would vary
by the initial abundance, but the pooled estimate is used for summarizing
the data.  The inability
to reproduce this quantity is likely to be reflected in misleading estimates of
demographic variance from \eqref{eq:var}.  To investigate this, 
Table \ref{tab:ecology:var} summarizes the demographic variance values
obtained from \eqref{eq:var}.  The table shows these variances for three species that correspond to the three quartiles of abundances at census interval 1 and follows these three species over time. Then \eqref{eq:var} is computed for the CTFS, UVC and MVC methods (with priors 1 and 2), and 
using either plug-in hyperparameter estimates or samples from the
posterior approximation, with the exception of UVC for which there is
no joint posterior estimate available.    
In general the CTFS method gives very different estimates of demographic
variance to the other summary statistic based approaches, which can be attributed partly to model misspecification, and warns of 
sensitivity of this variance to modelling assumptions.  
Also, we see that accounting
for posterior uncertainty using joint posterior samples makes a big difference
to the estimated variances for MVC, so that not being able to account for
this parameter uncertainty for the UVC method, which is univariate, is a limitation
of that approach.

%\begin{table}[ht]
%\centering
%\begin{tabular}{l|cccc}
%  \hline\hline
%census, species & CTFS & UVC & MVC (prior 1) & MVC (prior 2) \\ 
%  \hline
%$t=1, i=343$ & 2.99 & 5.27 & 9.03 & 9.35 \\
%  $t=2, i=343$ & 14.64 & 9.53 & 18.38 & 16.75 \\
%  $t=3, i=343$ & 11.49 & 12.23 & 22.29 & 20.78 \\
%  $t=4, i=343$ & 325.42 & 94.11 & 199.01 & 161.81 \\
%  $t=1, i=284$ & 21.55 & 41.81 & 78.28 & 77.97 \\
%  $t=2, i=284$ & 161.46 & 66.12 & 158.79 & 118.34 \\
%  $t=3, i=284$ & 64.15 & 80.11 & 146.41 & 129.94 \\
%  $t=4, i=284$ & 219.10 & 67.50 & 139.03 & 115.36 \\
%  $t=1, i=540$ & 130.62 & 280.56 & 561.32 & 542.58 \\
%  $t=2, i=540$ & 963.96 & 307.84 & 846.68 & 555.20 \\
%  $t=3, i=540$ & 210.79 & 286.17 & 526.98 & 456.48 \\
%  $t=4, i=540$ & 483.46 & 133.60 & 285.61 & 228.38 \\
%   \hline\hline
%\end{tabular}\caption{\footnotesize \emph{Ecology data.} Demographic variance for three selected species at each census interval.}\label{tab:ecology:var}
%\end{table}

\begin{table}[ht]
\centering\footnotesize
\begin{tabular}{l|ccccccc}
  \hline\hline
 & CTFS & CTFS(*) & UVC & MVC  & MVC  & MVC & MVC \\
census, species & & & &(prior 1) &(prior 1,*) &(prior 2) &(prior 2,*) \\
  \hline
$t=1, i=343$ & 2.99 & 3.01 & 5.27 & 4.79 & 9.04 & 5.28 & 9.35 \\
  $t=2, i=343$ & 14.67 & 14.83 & 9.50 & 10.47 & 18.39 & 9.23 & 16.76 \\
  $t=3, i=343$ & 11.51 & 11.04 & 12.14 & 13.53 & 22.30 & 12.63 & 20.81 \\
  $t=4, i=343$ & 324.01 & 337.00 & 94.55 & 119.09 & 199.10 & 93.91 & 161.86 \\
  $t=1, i=284$ & 21.51 & 21.42 & 41.87 & 40.87 & 78.34 & 42.60 & 77.88 \\
  $t=2, i=284$ & 162.21 & 165.09 & 65.84 & 91.51 & 158.74 & 65.82 & 118.32 \\
  $t=3, i=284$ & 63.82 & 61.07 & 80.37 & 93.63 & 146.56 & 84.96 & 129.79 \\
  $t=4, i=284$ & 216.96 & 225.90 & 67.27 & 82.85 & 139.28 & 67.83 & 115.25 \\
  $t=1, i=540$ & 130.98 & 129.41 & 279.88 & 285.79 & 560.29 & 292.14 & 542.48 \\
  $t=2, i=540$ & 966.81 & 983.22 & 304.97 & 491.29 & 846.51 & 309.97 & 555.47 \\
  $t=3, i=540$ & 210.95 & 201.31 & 293.08 & 349.71 & 526.95 & 301.96 & 456.39 \\
  $t=4, i=540$ & 483.89 & 501.47 & 134.64 & 169.71 & 285.72 & 134.39 & 228.24 \\

   \hline\hline
\end{tabular}\caption{\footnotesize \emph{Ecology data.} Demographic variance for three selected species at each census interval. Values for UVC have been computed based on the posterior mean estimate of $\psivec$. For CTFS and MVC we report results based on the respective point estimates, but also using samples from the joint posteriors (indicated with *).}\label{tab:ecology:var}
\end{table}

\section{Discussion}\label{sec:discussion}
This paper outlines a new scalable multivariate distributional regression
method, which has a number of unique features. 
It combines marginal distributions $G_1,\ldots,G_p$ with the
implicit copula $C_{1:n}$ of a multivariate regression model in which the regression
coefficients are integrated out. The flexible marginal specification ensures
that the approach exhibits good marginal calibration in uncertainty quantification. This proves important
in both an econometric prediction application, as well for LFI.
%The entire distribution of the response, and not just the conditional mean, varies with the covariates.  
In the copula construction we use a novel multivariate
extension of the horseshoe prior with attractive properties and consider two priors for the
cross-equation correlation matrix $\Sigma$. 

The equations at~\eqref{eq:linmod} share a common covariate vector $\xvec_i$ for each response variable. This may limit the usefulness of the method in some examples, although there are many applications with a common covariate across dimensions, including LFI. The method allows for sizable covariate vectors (e.g. $q=50$ and $q=70$ in our two applications) and large sample sizes (e.g. $n=17,250$ in one of our application). Estimation speed is also important when using 
distributional regression methods to conduct LFI. We achieve this here by
applying efficient variational inference methods to an augmented posterior, so 
that
our estimation approach is scalable. Finally, regularization is provided 
by our novel multivariate extension of the horseshoe prior. This has standard horseshoe priors for the coefficients from each equation as marginals, yet is a dependent prior in an analogous fashion as a g-prior, and may have applications
beyond that discussed here.

%Crucially, in our 
%Our method is presently limited to having 
%the same covariate vector for each
%variable.  It is most suited to the setting where this covariate vector
%is low-dimensional, since we consider flexible basis expansions to make
%the implicit copula flexible, and the number of basis terms
%required increases rapidly with the dimension.  A strength of our approach
%is that it can be used when the response vector is high-dimensional, and
%our proposed factor prior for the correlation matrix is particularly useful in
%this case.  Our novel multivariate horseshoe prior  
%may have applications beyond the multivariate regression copula model
%considered here.      
%\input{acknowl}
%\newpage
\appendix
\section{Priors for $\Sigma$}\label{app:A}
\subsection*{Prior~1: matrix logarithm}
Let $\text{vecl}(A)$ denote 
the half vectorization of the strictly lower triangular elements of a symmetric matrix $A$, 
then set $\upsilonvec=\text{vecl}(\log(\Sigma))$, where $\log(\Sigma)$ is the matrix
logarithm of $\Sigma$. \cite{ArcRei2021} show there is a one-to-one mapping 
between $\Sigma$ and $\upsilonvec$. To compute $\Sigma$ from $\upsilonvec$ these authors propose
the following recursive algorithm.
For a vector $\dvec\in \mathbb{R}^p$, 
write $A(\upsilonvec,\dvec)$ for the $p\times p$ symmetric matrix
with diagonal elements $\dvec$ and $\text{vecl}(A(\upsilonvec,\dvec))=\upsilonvec$.  There is a unique value $\dvec^*$ such that $\exp(A(\dvec^*,\upsilonvec))$ is a correlation matrix, where $\exp(A)$ is the matrix exponential
of $A$. To find $\dvec^*$, 
let $\dvec^{(0)}\in \mathbb{R}^p$ denote some initial guess for $\dvec$. Consider the recursion
$$\dvec^{(k+1)}=\dvec^{(k)}-\log \text{diag}(\exp(A(\upsilonvec,\dvec^{(k)}))),$$
where 
$\text{diag}(B)$ denotes the vector of diagonal entries
of $B$ and the logarithm is taken element-wise.
\cite{ArcRei2021} show that $\lim_{k\rightarrow\infty} \dvec^{(k)}=\dvec^*$ 
and that the algorithm converges exponentially fast.
Moreover, 
an expression for the Jacobian $\frac{\partial^{p(p-1)/2}}{\partial\upsilonvec}\text{vecl}(\Sigma) $ is given in Section~4.3 of \cite{ArcRei2021}, which is required when computing variational
inference with Prior~1.

We use a ridge prior $\upsilonvec\sim N(\bm{0},\sigma_\upsilon^2I_{p(p-1)/2})$ with
an  inverse gamma hyper-prior $\sigma_\upsilon^2\sim \mbox{IG}(a_\upsilon,b_\upsilon)$, with  $a_\upsilon=b_\upsilon=0.001$. 
%A proof that the resulting prior for $\Sigma$ is order invariant is given in Appendix~\ref{app:A1}.
With this prior, the copula parameters are $\varthetavec=\{\upsilonvec,\log(\sigma^2_\upsilon),\thetavec\}$.

\subsection*{Prior~2: factor re-parameterization}
Set $\Upsilon=GG^\top+D$, where $D=\mbox{diag}(d)$ and $G=\{g_{ij}\}_{i=1:p,j=1:K}$ is $p \times K$, with $K\ll p$, then a factor prior uses
the parameterization
\begin{equation*}
	\Sigma = \mbox{diag}(\Upsilon)^{-1/2} \Upsilon \mbox{diag}(\Upsilon)^{-1/2}\\
	= \widetilde G \widetilde G^\top +\widetilde D
\end{equation*}
%\textcolor{red}{why do we not need this rescaling if D=I?}
where 
\begin{eqnarray*}
	\mbox{diag}(\Upsilon)^{-1/2}&=&\mbox{diag}\left((d_1+\sum_{j=1}^{K} g_{1j}^2)^{-1/2},
	\ldots,(d_p+\sum_{j=1}^{K} g_{pj}^2)^{-1/2} \right)\,,\\
	\widetilde G &= &\mbox{diag}(\Upsilon)^{-1/2}G = \{\tilde g_{ij}\}_{i=1:p,j=1:K}\\
	\tilde g_{ij} & = &g_{ij}/\sqrt{d_i+\sum_{j=1}^K g_{ij}^2}\\
	\widetilde D &= &\mbox{diag}\left(\frac{d_1}{d_1+\sum_{j=1}^K g_{1j}^2},\ldots,
	\frac{d_p}{d_p+\sum_{j=1}^K g_{pj}^2}\right)
\end{eqnarray*}
Setting
$D=I$, the upper triangle of $G$ to zeros (i.e. $g_{ij}=0$ for $j>i$) and 
the leading diagonal elements of $G$ positive, identifies the parameters.
\cite{murray2013} point out that (approximately) uninformative priors for
$g_{ij}$ are 
informative for $\Sigma$, and 
suggest using a Generalized Double Pareto prior for each element $ j\leq i$ with density 
\[ 
\pi(g_{ij})=\frac{a_g}{2 b_g}\left(1+\frac{|g_{ij}|}{b_g}\right)^{-(a_g+1)}\,,
\] 
and $a_g=3$ and $b_g=1$, which is written GPD(3,1). 
With this prior, the real-valued parameters are $\gvec^0=(\mbox{vecl}(G),l_{11},\ldots,l_{KK})^\top$, where
the logarithm $l_{ii} = \log(g_{ii})$  for $i=1,\dots,K$, and the copula parameters are $\varthetavec=\{\gvec^0,\thetavec\}$. 

%\section{Variational inference approach}
%
%\subsection{Data augmentation}
%
%Here we show that integrating out $\betavec$ from the likelihood 
%\eqref{model2} in Section \ref{sec:copprocess} using the prior \eqref{eq:betaprior} gives
%a likelihood proportional to \eqref{model1}.  First, observe that
%\begin{align}
%  \int p(\yvec_{1:n};\xvec_{1:n},\betavec,\varthetavec)p(\betavec|\Sigma)\,d\betavec 
% & = \frac{\prod_{i=1}^n \prod_{j=1}^p g_j(y_{i,j})}{\prod_{i=1}^n \prod_{j=1}^p\phi(z_{i,j})}\int \phi_{np}(\zvec_{1:n};SX\betavec,S(\sigma\otimes I_n)S)p(\betavec|\Sigma)\,d\betavec. \label{marglike}
%\end{align}
%We can write
%$$\phi_{np}(\zvec_{1:n};SX\betavec,S(\Sigma\otimes I_n)S)
%=|S|^{-1}\phi_{np}(S^{-1}\zvec_{1:n},X\beta,\Sigma\otimes I_n),$$
%and then from the argument in Section \ref{sec:copprocess}
%\begin{align*}
%\int \phi_{np}(\zvec_{1:n};SX\betavec,S(\Sigma\otimes I_n)S)p(\betavec|\Sigma)\,d\betavec & = |S|^{-1}\int \phi_{np}(S^{-1}\zvec_{1:n};X\betavec,(\Sigma\otimes I_n))p(\betavec|\Sigma)\,d\betavec \\
% & = |S|^{-1}\phi_{np}(S^{-1}\zvec_{1:n};0,V).
%\end{align*}
%Writing $V=S^{-1}RS^{-1}$, we see that the last expression is
%$\phi_{np}(\zvec_{1:n};0,R)$.  Hence \eqref{marglike} is equal
%to \eqref{model1}.

\setlength{\bibsep}{0pt plus 0.3ex}
\bibliography{litliste}
%\input{supporting}
%\newpage
%\input{tabs}
%\input{figs}
\end{document}